\documentclass[12pt]{article}
\usepackage{axodraw}

\makeatletter
\newcommand\eqref[1]{(\ref{#1})}
\newcommand\Chref[1]{\expandafter\MakeUppercase\chaptername~\ref{#1}}
\newcommand\Secref[1]{\expandafter\MakeUppercase\secrefname~\ref{#1}}
\newcommand\Appref[1]{\expandafter\MakeUppercase\appendixname~\ref{#1}}
\newcommand\Figref[1]{\expandafter\MakeUppercase\figurename~\ref{#1}}
\newcommand\Tbref[1]{\expandafter\MakeUppercase\tablename~\ref{#1}}
   \let\secref\Secref  \let\appref\Appref
\let\figref\Figref  \let\tbref\Tbref
\newcommand\secrefname{Section}
\newcommand\II{I\kern-0.1em I}
\newcommand\nohyphens{\hyphenpenalty=\@M\exhyphenpenalty=\@M\relax}
\newcommand\hepth[1]{\texttt{hep-th/#1}}
\newcommand\npb[3]{Nucl.\ Phys.\ \textbf{B#1} (#2) #3}
\newcommand\plb[3]{Phys.\ Lett.\ \textbf{#1B} (#2) #3}
\newcommand\jhep[3]{JHEP \textbf{#1} (#2) #3}
\newcommand\cqg[3]{Class.\ Quant.\ Grav.\ \textbf{#1} (#2) #3}
\newcommand\ap[3]{Ann.\ Phys.\ \textbf{#1} (#2) #3}
\DeclareSymbolFont{AMSb}{U}{msb}{m}{n}
\DeclareSymbolFontAlphabet{\mathbb}{AMSb}
\DeclareSymbolFont{boldletters}{OML}{cmm}{b}{it}
\DeclareSymbolFontAlphabet{\mathbit}{boldletters}
\newcommand\al{\alpha}          
\newcommand\CB{\mathcal{B}}
\newcommand\CO{\mathcal{O}}

\newcommand\CT{\mathcal{T}}
\newcommand\Bb{\mathbit{b}}
\newcommand\Bc{\mathbit{c}}
\newcommand\Bf{\mathbit{f}}
\newcommand\Bh{\mathbit{h}}
\newcommand\BF{\mathbit{F}}
\newcommand\BQ{\mathbit{Q}}
\newcommand\BX{\mathbit{X}}
\DeclareMathSymbol{\Bal}{\mathord}{boldletters}{"0B}
\DeclareMathSymbol{\Bphi}{\mathord}{boldletters}{"1E}
\DeclareMathSymbol{\Bps}{\mathord}{boldletters}{"20}
\DeclareMathSymbol{\Bch}{\mathord}{boldletters}{"1F}
\DeclareMathSymbol{\Bom}{\mathord}{boldletters}{"21}
\newcommand\inv{^{\raise.15ex\hbox{${\scriptscriptstyle -}$}\kern-.05em 1}} 
\newcommand\ext[1][]{\mathop{\raisebox{.2ex}{$\textstyle\bigwedge$}^{#1}}}
\newcommand\del{\partial}
\newcommand\dd[1]{\frac\del{\del#1}}
\newcommand\one{\hbox{1\kern-.25em l}}                  
\newcommand\e{\mathrm{e}}                               
\newcommand\dint[2][]{\mathop{\mathalpha{\int#1}#2}}    
\newcommand\Bla{\Bigl\langle}
\newcommand\Bra{\Bigr\rangle}
\newcommand\Vev[1]{\Bigl\langle #1 \Bigr\rangle}
\newcommand\gerst[1]{\lbrace #1 \rbrace}
\newcommand\hoch[2][]{\ifnum#2=0 A\else\ifnum#2=1 \End(A)\else \Hom(A^{\otimes#2},A)\fi\fi ^{#1}}
\newcommand\rarr{\rightarrow}
\newcommand\darr{\downarrow}
\newcommand\labdarr[1]{\llap{$\scriptstyle #1$}\kern-3pt\downarrow}
\newcommand\labrarr[1]{\stackrel{#1}\rightarrow}
\newcommand\labsearr[1]{\searrow\kern-3pt\rlap{\kern-5pt\raisebox{5pt}{$\scriptstyle #1$}}}
\newcommand\set[1]{\mathbb{#1}}                         
\newcommand\C{\set{C}}                                  
\newcommand\R{\set{R}}
\newcommand\Z{\set{Z}}
\newcommand\group[1]{\mathop{\kern\z@\mathrm{#1}}\nolimits}     
\newcommand\U{\group{U}}                                
      
\newcommand\SO{\group{SO}}      
\newcommand\opname[1]{\mathop{\kern\z@\mathrm{#1}}\nolimits}    
\newcommand\Hom{\opname{Hom}}                           
\newcommand\End{\opname{End}}                           
\newcommand\Hoch{\opname{Hoch}}                         
\newcommand\Def{\opname{Def}}                           
\ifcase \@ptsize
\or 
\or 
\fi
  \divide\@tempdima\baselineskip
  \@tempcnta=\@tempdima
\renewcommand\section{\@startsection{section}{1}{\z@}%
                                    {-7ex \@plus -1ex \@minus -.2ex}%
                                    {2.5ex \@plus.2ex}%
                                    {\normalfont\large\scshape\centering}} 
\renewcommand\subsection{\@startsection{subsection}{2}{\z@}%
                                       {-5ex \@plus -1ex \@minus -.2ex}%
                                       {1.5ex \@plus.2ex}%
                                       {\normalfont\normalsize\scshape}}
\newcommand\ack{\section*{\ackname}}                                   
\newcommand\ackname{Acknowledgements}
\newcommand\sectionname{}

\renewcommand\@seccntformat[1]{\ignorespaces\csname #1name\endcsname\space%
  \csname the#1\endcsname.\quad}
\renewcommand\appendix{\par
  \setcounter{section}{0}%
  \setcounter{subsection}{0}%
  \renewcommand\thesection{\@Alph\c@section}%
  \renewcommand\sectionname{\appendixname}}
\long\def\@makecaption#1#2{%
  \vskip\abovecaptionskip
  \sbox\@tempboxa{\textsc{#1}:\space\slshape #2}%
  \ifdim \wd\@tempboxa >\hsize
    \textsc{#1}:\space\slshape #2\par
  \else
    \global \@minipagefalse
    \hb@xt@\hsize{\hfil\box\@tempboxa\hfil}%
  \fi
  \vskip\belowcaptionskip}
\def\eqnarray{%
   \stepcounter{equation}%
   \def\@currentlabel{\p@equation\theequation}%
   \global\@eqnswtrue
   \m@th
   \global\@eqcnt\z@
   \tabskip\@centering
   \let\\\@eqncr
   $$\everycr{}\halign to\displaywidth\bgroup
       \hskip\@centering$\displaystyle\tabskip\z@skip{##}$\@eqnsel
      &\global\@eqcnt\@ne$\;\hfil{##}$\hfil
      &\global\@eqcnt\tw@$\;\displaystyle{##}$\hfil\tabskip\@centering
      &\global\@eqcnt\thr@@ \hb@xt@\z@\bgroup\hss##\egroup
         \tabskip\z@skip
      \cr
}
\makeatother

\begin{document}

%
%
\thispagestyle{empty}

\begin{flushright}\scshape
RUNHETC-2001-04, UG-01-25\\
hep-th/0102201\\ 
Februari 2001
\end{flushright}
\vskip5mm

\begin{center}

{\LARGE\scshape
Deformations of Closed Strings and Topological Open Membranes
\par}
\vskip15mm

\textsc{Christiaan Hofman$^{1\dagger}$ \textnormal{and} Whee Ky Ma$^{2\ddagger}$}
\par\bigskip
{\itshape
  ${}^1$New High Energy Theory Center, Rutgers University,\\ 
        136 Frelinghuysen Road, Piscataway, NJ 08854, USA,
  \par\medskip
  ${}^2$Institute for Theoretical Physics, University of Groningen,\\
        Nijenborgh 4, 9747 AG Groningen, The Netherlands
}\par\bigskip
\texttt{${}^\dagger$hofman@physics.rutgers.edu, ${}^\ddagger$W.K.Ma@phys.rug.nl} 

\end{center}

\section*{Abstract}

We study deformations of topological closed strings. A well-known
example is the perturbation of a topological closed string by itself,
where the associative OPE product is deformed, and which is governed
by the WDVV equations. Our main interest will be closed strings that
arise as the boundary theory for topological open membranes, where the
boundary string is deformed by the bulk membrane operators. The main
example is the topological open membrane theory with a nonzero 3-form
field in the bulk. In this case the Lie bracket of the current
algebra is deformed, leading in general to a correction of the Jacobi
identity. We identify these deformations in terms of deformation
theory. To this end we describe the deformation of the algebraic
structure of the closed string, given by the BRST operator, the
associative product and the Lie bracket. Quite remarkably, we find
that there are three classes of deformations for the closed string,
two of which are exemplified by the WDVV theory and the topological
open membrane. The third class remains largely mysterious, as we have
no explicit example.

\newpage
\setcounter{page}{1}
%
%

\section{Introduction}
\label{sec:intro}

In recent years there has been much interest in string theory towards 
noncommutative geometry and noncommutative gauge theories. It was found 
at first that noncommutative gauge theories gave a natural description 
of M-theory in the presence of a NS $B$-field \cite{codo}, or more 
generally for D-branes in the presence of a $B$-field \cite{dohu,hove,scho,seiwit}. 
This noncommutative gauge theory on the D-branes can be understood 
as a description of the open string (field) theory in a decoupling limit. 
Rather than being a special situation in open string theory, noncommutativity 
seems to be quite generic, and is closely connected to the extended 
nature of strings. 

The noncommutative star product can be understood in terms of 
deformation quantisation: a deformation of function algebras 
starting from a Poisson bracket \cite{kon1,kon2}. 
This mathematical problem was solved recently by Tamarkin \cite{tam}, and by 
Kontsevich and Soibelman \cite{kon3}. The explicit solution found by 
Kontsevich can be understood quite elegantly in terms of the perturbation 
theory of a particular simple topological string \cite{cafe}. 
In our paper \cite{homaOS} we considered formal generalisations of 
these deformations for general topological open strings. 
The deformation quantisation problem is a special case 
of the more general problem of deformations of associative algebras. 
The results showed that the problems of deforming an associative 
algebra and string theory are intimately connected. This parallels the 
Deligne ``conjecture'' in mathematics (see e.g. \cite{kon2}), which 
states that the deformation theory of a ``1-algebra'' is a ``2-algebra''. 
In general $d$-algebras are intimately connected to $d$-dimensional 
(topological) field theories: they are defined in terms of 
(tree level) products for local operators in $d$ dimensions. 
A 1-algebra is simply an associative algebra. Indeed, we know that point 
particles are described by quantum mechanics, and operators in quantum 
mechanics form an associative algebra. On the other hand, 
string theories (2-dimensional quantum field theories) in general have the 
structure of a Gerstenhaber algebra -- an algebra consisting of a 
product and a Lie bracket \cite{wizwi,zwie,stasheff,kvz} -- which is the same 
thing as a 2-algebra \cite{kon2}. Hence the Deligne conjecture 
can be interpreted as stating that the deformation of a 
point particle theory is described by a string theory \cite{cafe,homaOS}. 
Indeed, in the case of noncommutative geometry, the boundary theory of
the open string, which is a gauge theory, is deformed to a
noncommutative gauge theory in the sense of Connes by turning on a
$B$-field, which is a closed string operator coupling to the bulk of
the string.

The Deligne ``conjecture'', which is now proven, can be generalised to higher 
dimensions \cite{kon2,tam2}: the deformation theory of a $d$-algebra is 
conjectured to have the structure of a $(d+1)$-algebra. A natural question 
from this point of view is therefore whether the deformation theory of 
2-dimensional (topological) field theories, or more generally closed string 
(field) theories, can be described by open membranes. 

Parallel to this is the question what the effect is of the 3-form
field on the closed string theory. Indeed the natural generalisation
of the 2-form coupling to the bulk of the string is the 3-form field
in the case of the open membrane. This 3-form field can be interpreted
either as the field strength of a 2-form gauge field -- which couples
to the boundary string as a gauge field -- or as the $C$-field in
M-theory, or as the 3-form RR field in type \II{A} string theory.
Attempts to describe the effect in terms of constrained canonical 
quantisation have been undertaken recently \cite{bebe,kasa,matshi}. In these
papers a noncommutative deformation of loop space was suggested. A
natural situation where the effect of a 3-form occurs is the M-theory
membrane ending on a M5-brane. This situation is particularly relevant
as it may provide more insight about the still mysterious M5-brane. 
The place to study the effects are various
decoupling limits of the M5-brane theories, in particular the $(2,0)$
little string theory \cite{lomosh,aha}, and the recently proposed OM
theory point \cite{om,bebe2}. In these situations the decoupled theories
one studies can be interpreted as closed string theories. Moreover,
they can be seen as the boundary of the supermembrane. The $C$-field
is a bulk membrane deformation. The effect of this $C$-field can
therefore be interpreted as a deformation of a closed string by an
open membrane. Related to this by double dimensional reduction is the
Type \II{A} situation of a D2-brane ending on a D4-brane, in a certain
decoupling limit \cite{om,bebe2}.

A deformation theory of closed strings, especially in the context of
topological string theory, was already studied about a decade ago
\cite{witeqn,dvv}. However, this deformation theory concerned the
deformation of closed strings by the closed string operators
themselves.

As deformations of closed strings can come either from closed strings
or from membranes, the question arises which of the two describes the
proper deformation theory of closed strings. In this paper we study
the general deformation complex of closed string theories. We show the
connection of the string theory correlation functions and their
deformations to the abstract deformation complex. We find that in
general deformations of closed strings cannot be described by a single
deformation complex. The paper is organised as follows.

In \secref{sec:tcs}, we discuss two-dimensional topological field
theories, whose correlation functions have the structure of a
Gerstenhaber algebra. Not restricting to physical operators leads to
algebras up to homotopy, defined by higher correlation functions. They
contain generalisations of both the associative and the Lie structure,
and are combined into a $G_\infty$ algebra. We make a concrete
proposal for the $A_\infty$ part of the algebra. 

In \secref{sec:wdvv} we review deformations of the closed string
algebra by inserted closed string operators. The associativity of the
deformed product is guaranteed by the WDVV equations. We will argue
that this goes through for the full $G_\infty$ algebra; it turns out
that only the $A_\infty$ structure is deformed. The multilinear maps
deforming the products are seen to form a structure of Gerstenhaber
algebra themselves. We show that this is the same algebra as the
underlying algebra of the deforming operators. The associativity in
first order of the deformed product corresponds to the BRST-closedness
of the deforming operator.

In \secref{sec:defcplx} we describe the general structure of
deformations of closed strings. This is governed by the Hochschild
complex, which contains all possible deformations of algebraic
operations. The Hochschild complex is an algebra by
itself, part of whose structure is induced by the algebraic structure
that is deformed. In the case of the open string, this structure is
determined by the (undeformed) open string theory. For the closed
string however, we find that the full structure induced by the
undeformed closed string cannot be used to define a consistent
deformation theory. One can only consistent deform a
substructure. This leads a priori to three different classes of
deformation theories, reflected in three different structures of
complexes; which one is valid depends of course on the specific model
under consideration.

In \secref{sec:cplxes}, we specify the classes of deformation
complexes. The deformation of closed strings by themselves studied in
\secref{sec:wdvv} turns out to have structure of one of these three.
The second class, related to deformations of the $L_\infty$ structure,
is described by a 3-dimensional theory. This leads us to suggest that
it can be understood in terms of topological open membrane theories,
where the boundary string is deformed by bulk membrane operators. For
the third deformation complex, which should be described by a
2-dimensional theory, we have no explicit realisation.

In \secref{sec:bdystr} we discuss topological open membranes in a
general setting. We try to describe the deformation theory of the
boundary string theory by the membrane bulk operators. Though we are
not able to prove all Ward identities in detail, due to our lack
of understanding 3-dimensional conformal field theories, we argue that
indeed the $L_\infty$ structure is deformed, and that the deformation
theory has the structure of the second class of deformation complexes.

In \secref{sec:tom} we describe an explicit example for the
topological open membrane (TOM), which was defined in \cite{js}: an
open membrane with only a WZ term, defined by a closed 3-form
field. The undeformed boundary string theory is the closed string
version of the Cattaneo-Felder model \cite{cafe}. The coupling of the
bulk membrane to the $C$-field indeed deforms the closed string Lie
bracket. We find that it induces a trilinear operation, which gives a
correction to the Jacobi identity of the bracket.

In \secref{sec:concl} we mention some possible extensions and
relations to physical models, such as OM theory, self-dual little
strings, and M5-branes. On the basis of the structure 
that we found in the open membrane, we speculate about consistent 
generalisations of interacting 2-form gauge theories, such as 
``non-abelian'' 2-forms. 

\section{Topological Closed Strings}
\label{sec:tcs}

Topological field theories are supplied with a BRST operator $Q$, an
anticommuting scalar, squaring to zero. For the theory to be
independent of the metric, the energy-momentum tensor $T$ should be
BRST-exact. As $T$ generates translations, this implies the existence
of an operator $G$ such that
\begin{equation}\label{commQG}
\{Q,G\} = d.
\end{equation}
For the bosonic string for example, this operator is given by the mode 
$b_{-1}$ of the antighost. 
The operator $G$ is fermionic too and should be a 1-form on the worldvolume. 
Furthermore, there is a conserved $\U(1)$ symmetry, whose conserved charge is 
called ghost number, such that the BRST operator $Q$ has ghost number 1, and the 
energy-momentum tensor, along with all physical operators, has ghost number zero. 
This implies that $G$ has ghost number $-1$.

Starting from any operator $\al\equiv \al^{(0)}$ that is a scalar on
the worldsheet, one can repeatedly use the operator $G$ to define
other operators, denoted $\al^{(p)}$, by the relation
$\al^{(p)}=\{G,\al^{(p-1)}\}$. They are called called descendants. As
$G$ is a 1-form, the descendant $\al^{(p)}$ is a $p$-form on the
worldsheet. Due to the anticommutation relations \eqref{commQG}, they
satisfy the descent equations, $Q\al^{(p+1)} = d\al^{(p)}$. Using
anticommuting coordinates $\theta^\mu$ on the worldsheet, one can
combine the operator $\al$ and its descendants into a ``superfield'',
$\Bal=\al+\theta\al^{(1)}+\frac12\theta^2\al^{(2)}$, where
contractions are suppressed in the notation. The condition for
physical or BRST-closed operators $\al$, \ $Q\al=0$, is now equivalent
to closedness of the superfield with respect to the full derivation
$Q+D$, where the superderivative operator $D=\theta^\mu\del_\mu$ is
introduced. We will assume that the scalar operator is BRST-closed,
unless stated otherwise.

For any operator and its descendants we can build corresponding 
observables by integrating them. The basic local observable is the 
evaluation of the operator in a point $x$, $\al(x)$. The descendants 
give rise to nonlocal observables $\int_{C_p}\al^{(p)}$, 
where in general $C_p$ is a $p$-cycle in the worldvolume. 
Note that the second descendant can be used to deform the action, 
$\delta S = \int_\Sigma\al^{(2)}$. The descent equations guarantee 
that these observables are BRST-closed and only depend on the homology 
class of the cycle $C_p$. For example, 
\begin{equation}
\al(x')-\al(x) = \int_x^{x'}d\al = \int_x^{x'}\{Q,G\}\al 
= Q\int_x^{x'}\al^{(1)}, 
\end{equation}
which decouples as it is BRST-exact. 

Next we discuss the correlation functions in the topological string theory. 
They can be identified with an 
algebraic structure on the operators in the closed string theory. 
For example the three-point functions determine a product structure. 
We now discuss the general structure of the algebra of closed string 
operators $\al_a$ at genus 0. 
There are two types of three-point functions. The most direct one involves 
just operators transforming as scalars, 
\begin{equation}
F_{abc} = \Vev{\al_a\al_b\al_c}. 
\end{equation} 
We assume that there is a special operator $\one$. Inserting it gives 
two-point functions $\eta_{ab}=F_{ab0}$, where the index $a=0$ 
denotes the special operator. It defines a metric on the space of worldsheet operators. 
Using the metric, we can raise and lower indices. This allows us to interpret 
the three-point functions as structure constants for a symmetric product on 
the space of operators, $\al_a\cdot\al_b = F^c{}_{ab}\al_c$. 
In this paper we will often denote this product by $m$. 
The operator $\one$ serves as a unit for this algebra. 
We can also construct correlators involving descendants. The natural three-point 
function is 
\begin{equation}
G_{abc} = \Bla \al_a \oint_{C}\!\al_b^{(1)} \al_c \Bra,
\end{equation}
where $C$ is a cycle enclosing the insertion point of $\al_c$ and not 
that of $\al_a$. Since we can contract the cycle, this is basically the only 
three-point function we can construct, except for adding top forms integrated 
over the worldvolume. It defines the structure constants of a graded antisymmetric product, 
called the bracket, 
$\{\al_a,\al_b\} = \oint_{C}\!\al_a^{(1)}\al_b=G^c{}_{ab}\al_c$. 
We will denote this bracket also by $b$. It plays an important role 
in the symmetry algebra of the string theory. Indeed, the first descendant 
is a current, which acts in this way.

The operations defined by the three-point functions satisfy several
well-known relations. These relations, which we will discuss and
generalise below, follow from factorisation of the higher correlation
functions. First of all, the product $m$ is associative. The bracket
$b$ satisfies the Jacobi identity, therefore it is a Lie bracket as
expected. The associative product $m$ and the Lie bracket $b$ also
satisfy a mutual compatibility, which is similar to the one found for
a Poisson algebra. Together, they form an algebra which is thus much
like a Poisson algebra. The only difference is that the bracket $b$
has ghost number $-1$, due to the descendant theory. The resulting
structure is called a \emph{Gerstenhaber algebra} ($G$
algebra), see \appref{app:hom}.\footnote{Gerstenhaber algebras might be 
more familiar to physicists as substructures of Batalin-Vilkovisky (BV) 
algebras, after forgetting the BV operator. In fact, in all known cases 
closed string theories have the full structure of a BV algebra. It is 
however not known to us at this point that this necessarily should be the case.}

This structure of a $G$ algebra is naturally connected to the topology
of 2-dimensional surfaces. If we consider products, we need to insert
two operators corresponding to the ``in''-state. We start by putting
an operator on a point; this corresponds to a puncture in the
plane.\footnote{The boundary at infinity corresponds to the
``out''-state.}  The topology of this punctured plane $\Sigma_1$
remaining for the second operator has two generators: a point, the
generator of $H_0(\Sigma_1)=\Z$, and a circle enclosing the puncture,
corresponding to $H_1(\Sigma_1)=\Z$. These two generators of the
topology naturally correspond to the two bilinear operations in the
algebra. The fact that the second nontrivial homology is concentrated
in degree one corresponds through the descent equations to the fact
that the bracket has degree $-1$. 

The above is precisely the picture arising from the operad of little 
discs \cite{kon2}. These are formal structures of discs with holes and 
gluing relations, which is indeed closely related to formal definitions 
of topological strings. $G$ algebras arise as algebras over the 
(singular) homology of this operad, and were dubbed 2-algebras in 
this context.

\subsection{Higher Correlation Functions}

Off-shell the Ward identities giving algebraic relations such as the associativity 
and the Jacobi identity will acquire mild corrections involving the BRST operator. 
These will involve higher order correlation functions, and the corresponding 
multilinear operations generate more involved algebraic structures called homotopy 
algebras. For example, the lowest order correction to the associativity of the 
product involves a trilinear operation. More generally, the associative 
algebra becomes a homotopy associative or $A_\infty$ algebra, see 
e.g.\ \cite{stas,getzjon1,getzjon2,zwieoc,kon4}. 
Also the Jacobi identity for the bracket will get higher order corrections. 
The corresponding multilinear maps generate a homotopy Lie or $L_\infty$ algebra, 
see e.g.\ \cite{wizwi,zwie,stasheff,ksv,kon2}. Definitions of $A_\infty$ 
and $L_\infty$ are given in \appref{app:hom}, and will also 
be discussed in more detail later.  

As the product and the Lie bracket combine into a Gerstenhaber algebra, 
it is expected that the $A_\infty$ and $L_\infty$ structure are part of 
a homotopy Gerstenhaber or $G_\infty$ structure \cite{ksv,stasheff}. 
Several proposals for this structure appeared in the literature. 
An operad definition of $G_\infty$ was discussed in \cite{liz,kvz}, in 
connection to (topological) strings. Another version called $B_\infty$ 
was discussed in \cite{getzjon2}. In \cite{tamtsy,tam} 
a related but more general algebraic definition for $G_\infty$ 
appeared in the context of deformation quantisation and formality. 
Here we will restrict to a discussion of the $A_\infty$ and 
$L_\infty$ subalgebras. 

Once we take the operators on-shell, the higher correlators remain
in general. They still satisfy homotopy algebra relations, 
though the $G$ algebra decouples in the sense that it will be a 
genuine subalgebra. This makes it possible to discuss the on-shell description 
purely in terms of this $G$ algebra structure, as is often done in the literature. 
However, this is not natural in the context of deformations, as in general 
the BRST operator deforms, as noted for example, in \cite{verl}. Furthermore, 
higher operations appear naturally in deformation theory, and also turn out 
to play a crucial role in deformations of string theory, as we will see later, 
and these appear most naturally off-shell as explained above. It is very 
hard to to give a complete off-shell definition, and structures are not defined 
canonically, but rather depend crucially on the insertion points. 
The definitions we will give below are preliminary, and are strictly speaking 
only well-defined on-shell. 

The basis of this structure -- the BRST operator, the bracket, and the 
product -- was discussed above. The higher structure constants of the $L_\infty$ 
algebra are defined by the following correlation functions
\begin{equation}\label{linfcorr}
G_{a_0a_1\ldots a_n} = 
 \Vev{ \al_{a_0}\oint\!\al_{a_1}^{(1)}\al_{a_2}\int\!\al_{a_3}^{(2)}\cdots\int\!\al_{a_n}^{(2)} }
 =: \Vev{ \al_{a_0}\; b_n(\al_{a_1},\cdots,\al_{a_n})},
\end{equation}
where the last equality defines the higher multilinear brackets $b_n$ 
of the $L_\infty$ algebra. 
It can indeed be shown, using the Ward identities and factorisation, that 
the corresponding multilinear maps satisfy the $L_\infty$ relations. 
They are the multilinear string products of 
\cite{wizwi,zwie} expressed in local coordinates on the moduli space (at genus 0), 
generalised to topological strings by replacing $b$ with $G$. The 
proof of the $L_\infty$ relations given in \cite{zwie} also applies here. 
Furthermore, the Ward identities for the spin-2 field $G$ assure the graded 
antisymmetry of these structure constants. 

The $A_\infty$ algebra is a bit more involved. As far as we know, 
the full $A_\infty$ structure has not been studied in the literature, at 
least we are not aware of any explicit formulas. We propose the following 
definition of the structure constants for the higher products $m_n$ 
\begin{eqnarray}\label{ainfcorr}
F_{a_0a_1\ldots a_n} &=&  
 (-1)^{(n-2)g_1+(n-3)g_2+\ldots+g_{n-2}}\Vev{ \al_{a_0}\al_{a_1}
\int_1^n\!\al_{a_2}^{(1)}\int_2^n\!\al_{a_3}^{(1)}\cdots\int_{n-2}^n\!\al_{a_{n-1}}^{(1)} 
\al_{a_n}}
\nonumber\\
 &=:& \Vev{ \al_{a_0}\; m_n(\al_{a_1},\cdots,\al_{a_n})},
\end{eqnarray}
where $g_k=|\alpha_{a_k}|$ denotes the (ghost) grading of the operator 
$\alpha_{a_k}$.\footnote{The signs are added because we want to contribute 
signs coming from the descendants to the operation $m_n$.} 
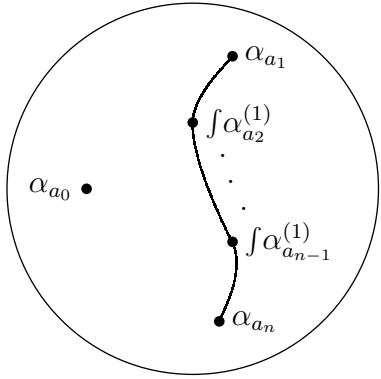
\begin{figure}[hr]
\begin{center}
\begin{picture}(150,150)(-5,-5)
\CArc(70,70)(70,0,360)
\Vertex(30,70)2  \Text(25,70)[r]{$\al_{a_0}$}
\Vertex(85,120)2 \Text(90,120)[l]{$\al_{a_1}$}
\Vertex(80,20)2  \Text(85,20)[l]{$\al_{a_n}$}
\Vertex(70,95)2  \Text(75,95)[l]{$\int\!\al_{a_2}^{(1)}$}
\Vertex(85,50)2  \Text(90,50)[l]{$\int\!\al_{a_{n-1}}^{(1)}$}
\Text(80,83)[l]{.}  \Text(83,73)[l]{.} \Text(88,63)[l]{.}
\qbezier(85,120)(70,105)(70,95)
\qbezier(70,95)(70,80)(85,50)
\qbezier(85,50)(90,40)(80,20)
\end{picture}
\end{center}
\caption{The correlation functions on the sphere defining the $A_\infty$ 
structure constants. The first descendants are integrated along the indicated 
path in path order.}
\label{ainfty}
\end{figure}

They are depicted in \figref{ainfty}. 
They involve a chain of path-ordered integrals along a path connecting the 
insertion points of $\al_{a_1}$ and $\al_{a_n}$. These structure constants 
are not symmetric for $n\geq3$. They do have however certain symmetry properties 
due to the Ward identities, at least on-shell. For example, independence of 
the choice of path  on-shell.

Let us motivate this proposal. We start with the trilinear product. The relations in 
the $A_\infty$ algebra relate this to an off-shell correction to the associativity 
of the product. 
Usually, one proves associativity by considering the factorisation of the four-point 
function $\Vev{\al_a\al_b\al_c\al_d}$ into two three-point functions. 
Consistency of the factorisation in the $s$-channel and the $t$-channel then 
gives associativity. The factorisation however is corrected once we allow off-shell 
operators. To find the correct formula, we write the difference of the $s$-channel 
and the $t$-channel factorisation as boundary terms of an integral 
\begin{equation}
\Vev{ \al_{a_0}\al_{a_1}\int_1^3\!d\al_{a_2}\al_{a_3} } =
 \Vev{ \al_{a_0}\al_{a_1}\al^b } \Vev{ \al_b\al_{a_2}\al_{a_3} }
 -\Vev{ \al_{a_0}\al^b\al_{a_3} } \Vev{ \al_b\al_{a_1}\al_{a_2} }
\end{equation}
We can use the descent equations to write the total derivative as 
$d\al_{a_2}=Q\al_{a_2}^{(1)}+(Q\al_{a_2})^{(1)}$, and move the 
BRST operator in the first term to the other operators. We find the relation 
\begin{eqnarray}\label{assoc}
&&\kern-3em 
\Vev{ \al_{a_0}\al_{a_1}\al^b } \Vev{ \al_b\al_{a_2}\al_{a_3} }
 -\Vev{ \al_{a_0}\al^b\al_{a_3} } \Vev{ \al_b\al_{a_1}\al_{a_2} } 
\nonumber\\
 &=& \Vev{ \al_{a_0}\al_{a_1}\int_1^3\!(Q\al_{a_2})^{(1)}\al_{a_3} } 
 +(-1)^{g_0+g_1}\Vev{ Q\al_{a_0}\al_{a_1}\int_1^3\!\al_{a_2}^{(1)}\al_{a_3} } \\
 && +(-1)^{g_1}\Vev{ \al_{a_0}Q\al_{a_1}\int_1^3\!\al_{a_2}^{(1)}\al_{a_3} } 
 -(-1)^{g_2}\Vev{ \al_{a_0}\al_{a_1}\int_1^3\!\al_{a_2}^{(1)}Q\al_{a_3} } 
\nonumber
\end{eqnarray}
If all operators are on-shell, this indeed proves associativity. 
However, off-shell we find corrections from the right-hand side. 
These corrections have precisely the form of the 
four-point functions for the $A_\infty$ structure we proposed. 
We can interpret the factorised correlation functions in terms of 
compositions of the multilinear maps forming this $A_\infty$ structure. 
Explicitly, we can write \eqref{assoc} as 
\begin{eqnarray}\label{assoc2}
&&\kern-3em 
m(m(\al_{a_1},\al_{a_2}),\al_{a_3}) - m(\al_{a_1},m(\al_{a_2},\al_{a_3})) \nonumber\\ 
 &=& - Q(m_3(\al_{a_1},\al_{a_2},\al_{a_3}))  
 - m_3(Q\al_{a_1},\al_{a_2},\al_{a_3}) \\
&& - (-1)^{g_1}m_3(\al_{a_1},Q\al_{a_2},\al_{a_3})
 - (-1)^{g_1+g_2}m_3(\al_{a_1},\al_{a_2},Q\al_{a_3})
\nonumber
\end{eqnarray}
We can formally write this relation in the form $m\circ m = - Q\circ m_3 - m_3\circ Q$,
where $\circ$ is a certain composition of multilinear maps. Note that this 
composition involves summing over different permutations. The precise 
definition of this composition will be discussed in more detail later, 
but can of course be read off from the factorisation in general. 
This is the correction of the associativity one finds in an 
$A_\infty$ algebra. 

A similar analysis can be performed for the higher products. Though we have 
not carried out the complete analysis, we will give the general idea of a proof. 
Commuting a BRST operator through the formula for the higher product $m_n$, 
one similarly finds boundary terms. These can be viewed as the chain 
of $n-2$ ordered integrals being broken up into two chains of length 
$n_1-2$ and $n_2-2$, where $n_1+n_2=n+1$. Note that we need $n_1,n_2\geq2$. 
These boundary terms factorise. This gives a relation of the form 
\begin{equation}
Q\circ m_n\pm m_n\circ Q = \sum_{n_1+n_2=n+1} (\pm)  m_{n_1}\circ m_{n_2},
\end{equation}
where the signs are  determined by the various degrees. This reproduces 
the $A_\infty$ relations, as will be discussed in more detail later.

\section{Deformed Correlators and Algebraic Structure}
\label{sec:wdvv}

In this section we discuss deformations of the correlation functions,
and therefore of the algebra of the topological closed string theory,
by inserting extra closed string operators. The WDVV equations show
that these correlators can indeed be interpreted as deformations of
the correlation functions. We also discuss how the Gerstenhaber
structure of the deforming operators is translated into the
Gerstenhaber structure of the multilinear maps.

\subsection{WDVV Equations}

We can define higher correlators by inserting integrated second 
descendants. The closed string Ward identities for $G$ assure that these 
correlators are symmetric in the closed string indices \cite{witeqn,dvv}. 
These relations are known as the WDVV equations. They imply an integrability 
of the correlation functions: there must exist a function $F(t)$ of formal 
parameters $t^a$, such that the higher correlators can be found by 
differentiating this function. For example, the three-point functions are given 
by $F_{abc}=\del_a\del_b\del_cF(t)$. Setting $t=0$ in this relation gives back 
the original structure constants. However, this equation is valid for nonzero 
$t$ as well, if we define the deformed three-point functions by formally 
exponentiating a deformation $\int\!\al^{(2)}$, 
\begin{equation}\label{deform}
F_{abc}(t) = 
\Vev{ \al_a\al_b\al_c\,\e^{\,t^d\!\!\int\!\al_d^{(2)}}},
\end{equation}
where the exponentiated second descendant can be identified with 
a deformation of the action functional. 
It shows that indeed the insertions of closed string operators deform the 
closed string algebra, yielding a deformed $A_\infty$ algebra. 


We like to describe the WDVV equations in the context of deformation theory. 
To facilitate this relation, we will distinguish in the notation between the 
operators in the algebra and the operators that are used to deform it. We
use the notation $\al_a$ for the operators in the algebra $A$ we want to 
deform and $\phi_i$ for the deforming operators, although for now they are 
taken from the same algebra.\footnote{More generally, we could take for the 
algebras any algebraically closed subalgebras.}
Starting from our proposal \eqref{ainfcorr} for the $A_\infty$ algebra, 
we then write for the deformed higher-point functions 
\begin{equation}\label{deformA}
\Phi_{ia_0a_1\ldots a_n} = (-1)^{(n-2)g_1+(n-3)g_2+\ldots+g_{n-2}}
\Vev{ \al_{a_0}\int\!\phi_i^{(2)}\al_{a_1}
 \int\!\al_{a_2}^{(1)}\cdots\int\!\al_{a_{n-1}}^{(1)}\al_{a_n} }.
\end{equation}
As for the undeformed $A_\infty$ structure, these definitions are well-defined and 
path independent if the operators $\alpha_a$ are taken on-shell, but we expect some 
generalisation off-shell. 
Upon introducing more deforming operators $\int\!\phi_j^{(2)}$ etc, the 
WDVV equations amount to symmetry with respect to all
deforming operators. 

The reason that we chose the deformations of correlation functions for the 
$A_\infty$ algebra rather than the correlation functions \eqref{linfcorr} 
defining the $L_\infty$ algebra is that the latter are not deformed on-shell, 
as can easily be seen from the Ward identities of $G$. 

We will interpret the  $\Phi_{ia_0\ldots a_n}$ in terms of a multilinear map 
$\Phi_i:A^{\otimes n}\rightarrow A$, through the following definition 
\begin{equation}\label{mixed}
\Phi_{ia_0a_1 \dots a_n} = \Vev{ \al_{a_0} \;\Phi_i(\al_{a_1}, \dots, \al_{a_n}) }.
\end{equation}
These maps are the infinitesimal deformations of the $A_\infty$ 
algebra structure constants. We will sometimes write $\Phi_i=\Phi(\phi_i)$, 
to emphasise the relation with the deforming operator. Note that any $\phi_i$ 
corresponds to an infinite set of maps, one for any order $n$. 

Let us now examine the deformation of the $A_\infty$ structure more closely. 
The first-order deformations are simply given by inserting an extra integrated 
second descendant. Using the Ward identity for $G$ we can also write the 
corresponding deformed correlator
\eqref{deformA} as 
\begin{equation}\label{higherA}
\Phi_{ia_0\ldots a_n} = (-1)^{(n-1)g_1+\ldots+g_{n-1}}
 \Vev{ \al_{a_0}\phi_i\int\!\al_{a_1}^{(1)}\cdots\int\!\al_{a_n}^{(1)} }. 
\end{equation}
The proof is almost the same as the corresponding one for the open string case 
in \cite{homaOS}. This formula has the advantage that it also applies 
to the deformation $n\leq1$. 
In particular, for $n=1$, it should give the deformation of the linear 
map $m_1=Q$ in the $A_\infty$ algebra. 
For $n=1$, we know we can also write the correlation function 
in the form 
\begin{equation}
\Phi_{iab} =  \Vev{ \al_{a}\oint\!\phi_i^{(1)}\al_{b} }.
\end{equation}
This is exactly the deformation of the BRST operator \cite{verl}. 

\subsection{Structure of the Deformation Maps}

We want to see the relation between the maps defined by the deformation 
and abstract deformation theory. To make the notation not too cumbersome, 
we will assume that the undeformed algebra is a genuine differential associative 
algebra. That is, the higher products $m_n$ for $n\geq3$ are zero. 
For later reference, let us first 
look at the degree of the operators. For any of the operators 
$\al_a$ we write its ghost number as $g_a$. The ghost number 
of its dual operator $\al^a$ (with respect to the metric defined 
by the two-point functions) is written $g^a$. If $\Delta$ is the ghost 
anomaly, this is given by $g^a=\Delta-g_a$. For the ghost number of the 
deforming operator we write $g_\phi$. When we consider the corresponding 
map $\Phi$ of order $n$, it has an internal ghost degree given by 
$g_\Phi=g^{a_0}-\sum_{k=1}^n g_{a_k}$. This ghost number is 
such that the total ghost number in \eqref{mixed} adds up to $\Delta$. 
Due to ghost number conservation, there is now a relation 
between the ghost number of the deforming operator $\phi$ and the 
ghost number of the corresponding map, given by 
the corresponding
\begin{equation}\label{ghostWDVV}
g_\phi = g^{a_0}-\sum_{k=1}^n(g_{a_k}-1) = g_\Phi+n. 
\end{equation}
The shift in the degree equals the order of the map. 
This shift is due to the descendants that appear in the correlation functions. 

The deforming operators $\phi_i$ form an algebra, as they are the
closed string operators themselves. However, their identification with
multilinear maps also gives them an algebraic structure. The algebraic
structure of the operators should translate into algebraic operations
on these maps. This relation will be crucial in connection with
deformation theory.

We start with the action of the BRST operator $Q$. In order to 
see this action, we need to consider a deforming operator 
$\phi$ not necessarily on-shell. 
Then there is the following relation 
\begin{eqnarray}
\Vev{ \al_a\int\!(Q\phi_i)^{(2)}\al_b\al_c } &=& 
 - \Vev{ \al_a\oint\!\phi_i^{(1)}\al^e}\Vev{\al_e\al_b\al_c } 
 + \Vev{ \al_a\al^e\al_c}\Vev{\al_e\oint\!\phi_i^{(1)}\al_b }, 
\nonumber\\ 
&& + (-1)^{(g_i-1)g_b}\Vev{ \al_a\al_b\al^e}\Vev{\al_e\oint\!\phi_i^{(1)}\al_c }, 
\end{eqnarray}
where the $\al$ operators are taken on-shell. 
For $\phi$ on-shell, the left-hand side is zero, and we can interpret the 
right-hand side as a deformed Leibniz rule. 
Similarly we find for the four-point function: 
\begin{eqnarray}
(-1)^{g_1}\Vev{\al_{a_0}\int\!(Q\phi_i)^{(2)}\al_{a_1}\int_1^3\!\al_{a_2}^{(1)}\al_{a_3}} 
 &=& 
\Vev{\al_{a_0}\int\!\phi_i^{(2)}\al_{a_1}\al^b}\Vev{\al_b\al_{a_2}\al_{a_3}} 
\nonumber\\ &&
 - \Vev{\al_{a_0}\int\!\phi_i^{(2)}\al^b\al_{a_3}}\Vev{\al_b\al_{a_1}\al_{a_2}} 
\\ &&
 + \Vev{\al_{a_0}\al^b\al_{a_3}}\Vev{\al_b\int\!\phi_i^{(2)}\al_{a_1}\al_{a_2}} 
\nonumber\\ &&
 + (-1)^{g_ig_1}
\Vev{\al_{a_0}\al_{a_1}\al^b}\Vev{\al_b\int\!\phi_i^{(2)}\al_{a_2}\al_{a_3}}.
\nonumber
\end{eqnarray}
This equality shows  that if 
$\phi_i$ is on-shell, that is $Q\phi_i=0$, the deformed product is associative, 
at least to first order in the deformation. Thus the BRST operator corresponds 
to the first-order deformed associator. 

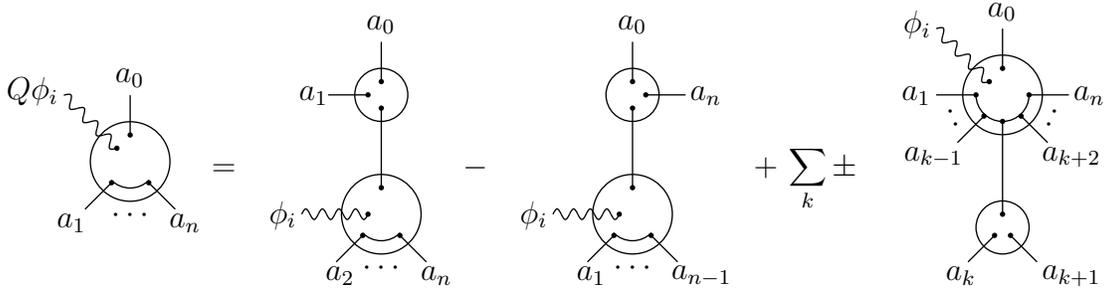
\begin{figure}[t]
\begin{center}
\begin{picture}(415,120)
\put(-5,20){
  \begin{picture}(80,80)(0,10)
    \CArc(50,45)(15,0,360)
    \Photon(25,70)(45,50)24 \Vertex(45,50)1 \Text(23,72)[r]{$Q\phi_i$}
    \Line(33,27)(43,37) \Vertex(43,37)1 \Text(28,22)[m]{$a_1$}
    \Line(67,27)(57,37) \Vertex(57,37)1 \Text(65,22)[l]{$a_n$}
    \Line(50,55)(50,70) \Vertex(50,55)1 \Text(50,77)[m]{$a_0$}
    \CArc(50,45)(10,225,315) \Text(50,25)[m]{$\cdots$}
  \end{picture}}
\put(80,50){$=$}
\put(100,0){
  \begin{picture}(80,120)(0,20)
    \CArc(40,100)(10,0,360)
    \Line(40,105)(40,120) \Vertex(40,105)1  \Text(40,127)[m]{$a_0$}
    \Line(35,100)(20,100) \Vertex(35,100)1  \Text(20,100)[r]{$a_1$}
    \CArc(40,55)(15,0,360)
    \Line(40,65)(40,95) \Vertex(40,65)1 \Vertex(40,95)1 
    \Photon(10,55)(35,55)24 \Vertex(35,55)1 \Text(8,55)[r]{$\phi_i$}
    \Line(23,37)(33,47) \Vertex(33,47)1 \Text(30,32)[r]{$a_{2}$}
    \Line(57,37)(47,47) \Vertex(47,47)1 \Text(55,32)[l]{$a_{n}$}
    \CArc(40,55)(10,225,315) \Text(40,35)[m]{$\cdots$}
  \end{picture}}
\put(175,50){$-$}
\put(195,0){
  \begin{picture}(80,120)(0,20)
    \CArc(40,100)(10,0,360)
    \Line(40,105)(40,120) \Vertex(40,105)1  \Text(40,127)[m]{$a_0$}
    \Line(45,100)(60,100) \Vertex(45,100)1  \Text(62,100)[l]{$a_n$}
    \CArc(40,55)(15,0,360)
    \Line(40,65)(40,95) \Vertex(40,65)1 \Vertex(40,95)1 
    \Photon(10,55)(35,55)24 \Vertex(35,55)1 \Text(8,55)[r]{$\phi_i$}
    \Line(23,37)(33,47) \Vertex(33,47)1 \Text(30,32)[r]{$a_{1}$}
    \Line(57,37)(47,47) \Vertex(47,47)1 \Text(55,32)[l]{$a_{n-1}$}
    \CArc(40,55)(10,225,315) \Text(40,35)[m]{$\cdots$}
  \end{picture}}
\put(285,50){$\displaystyle +\,\sum_k\pm$}
\put(335,0){
  \begin{picture}(80,120)(0,20)
    \CArc(40,100)(15,0,360)
    \Photon(15,125)(35,105)24 \Vertex(35,105)1 \Text(13,127)[r]{$\phi_i$}
    \CArc(40,100)(10,180,360)
    \Line(40,110)(40,125) \Vertex(40,110)1  \Text(40,132)[m]{$a_0$}
    \Line(15,100)(30,100) \Vertex(30,100)1  \Text(13,100)[r]{$a_1$}
    \Line(65,100)(50,100) \Vertex(50,100)1  \Text(67,100)[l]{$a_n$}
    \Line(23,82)(33,92)   \Vertex(33,92)1   \Text(25,77)[r]{$a_{k-1}$}
     \Text(22,94)[r]{$\cdot$} \Text(24,89)[r]{$\cdot$}
     \Text(58,94)[l]{$\cdot$} \Text(56,89)[l]{$\cdot$}
    \Line(57,82)(47,92)   \Vertex(47,92)1   \Text(55,77)[l]{$a_{k+2}$}
    \CArc(40,50)(10,0,360)
    \Line(40,55)(40,90) \Vertex(40,55)1 \Vertex(40,90)1 
    \Line(27,37)(37,47) \Vertex(37,47)1 \Text(30,32)[r]{$a_{k}$}
    \Line(53,37)(43,47) \Vertex(43,47)1 \Text(55,32)[l]{$a_{k+1}$}
  \end{picture}}
\end{picture}
\end{center}
\caption{Factorisation expressing the action of the BRST operator.}
\label{delmfact}
\end{figure}

There is a generalisation to higher correlators, which is depicted in 
\figref{delmfact}. 
This relation can be stated as 
\begin{equation}\label{WDVVQ}
\Phi(Q\phi_i)=m\circ\Phi(\phi_i)\pm\Phi(\phi_i)\circ m.
\end{equation}
This shows that $Q$ is represented on the algebra of 
maps on the cohomology by $\Phi\circ m + m\circ\Phi$, where $m$ is the 
product of the closed string algebra. If the operators $\al$ are not on-shell either, 
we get corrections from the BRST operator acting on the various 
$\al$'s. $Q\phi$ then corresponds
to $Q\circ\Phi_i\mp\Phi_i\circ Q+m\circ\Phi_i\pm\Phi_i\circ m$, 
where the first terms can be expanded as 
\begin{equation}
(Q\circ \Phi_i\mp\Phi_i\circ Q)(\al_{a_1},\ldots,\al_{a_n}) = Q\Phi_i(\al_{a_1},\ldots,\al_{a_n})
-\sum_i \pm\Phi_i(\al_{a_1},\ldots,Q\al_{a_i},\ldots,\al_{a_n}).
\end{equation}

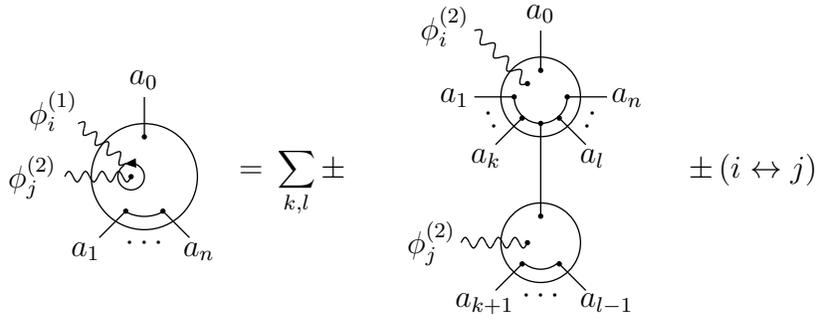
\begin{figure}
\begin{center}
\begin{picture}(310,130)
\put(0,20){
  \begin{picture}(75,90)(0,10)
    \CArc(50,50)(20,0,360)
    \Photon(25,70)(43,55)24
    \LongArrowArc(45,50)(5,120,120)  \Text(25,75)[r]{$\phi_i^{(1)}$}
    \Photon(20,50)(45,50)24  \Vertex(45,50)1  \Text(17,50)[r]{$\phi_j^{(2)}$}
    \Line(33,27)(43,37) \Vertex(43,37)1 \Text(28,22)[m]{$a_1$}
    \Line(67,27)(57,37) \Vertex(57,37)1 \Text(65,22)[l]{$a_n$}
    \Line(50,65)(50,80) \Vertex(50,65)1 \Text(50,87)[m]{$a_0$}
    \CArc(50,50)(15,240,300) \Text(50,25)[m]{$\cdots$}
  \end{picture}}
\put(90,60){$\displaystyle =\, \sum_{k,l}\pm$}
\put(160,0){
  \begin{picture}(80,130)(0,10)
    \CArc(40,100)(15,0,360)
    \Photon(15,125)(35,105)24 \Vertex(35,105)1 \Text(13,127)[r]{$\phi_i^{(2)}$}
    \CArc(40,100)(10,180,360)
    \Line(40,110)(40,125) \Vertex(40,110)1  \Text(40,132)[m]{$a_0$}
    \Line(15,100)(30,100) \Vertex(30,100)1  \Text(13,100)[r]{$a_1$}
    \Line(65,100)(50,100) \Vertex(50,100)1  \Text(67,100)[l]{$a_n$}
    \Line(23,82)(33,92)   \Vertex(33,92)1   \Text(25,77)[r]{$a_{k}$}
     \Text(22,94)[r]{$\cdot$} \Text(24,89)[r]{$\cdot$}
     \Text(58,94)[l]{$\cdot$} \Text(56,89)[l]{$\cdot$}
    \Line(57,82)(47,92)   \Vertex(47,92)1   \Text(55,77)[l]{$a_{l}$}
    \CArc(40,45)(15,0,360)
    \Line(40,55)(40,90) \Vertex(40,55)1 \Vertex(40,90)1 
    \Photon(10,45)(35,45)24 \Vertex(35,45)1 \Text(8,45)[r]{$\phi_j^{(2)}$}
    \Line(23,27)(33,37) \Vertex(33,37)1 \Text(30,22)[r]{$a_{k+1}$}
    \Line(57,27)(47,37) \Vertex(47,37)1 \Text(55,22)[l]{$a_{l-1}$}
    \CArc(40,45)(10,225,315) \Text(40,25)[m]{$\cdots$}
  \end{picture}}
\put(260,60){$\pm\, (i\leftrightarrow j)$}
\end{picture}
\end{center}
\caption{Factorisation for the bracket.}
\label{bfact}
\end{figure}

There are also relations between the products and the brackets of deforming 
operators on the one hand and of factorised correlation functions on the other hand. 
For these we will be a bit less precise, and only consider the general form. 
To study them, we have to look at the second-order terms, including two deforming 
operators. Again we interpret the factorised correlation functions as algebraic 
operations on the maps $\Phi_i$ and $\Phi_j$. For the bracket we study
\begin{equation}
\Vev{\int\!(Q\phi_i)^{(2)}\int\!\phi_j^{(2)}\al_{a_0}\al_{a_1}
 \int\!\al_{a_2}^{(1)}\cdots\int\!\al_{a_{n-1}}^{(1)}\al_{a_n}} = 0. 
\end{equation}
Passing the $Q$ through the descendants gives at one side a boundary term 
for $\phi_i^{(1)}$ coming close to $\phi_j$, which is the map corresponding 
to $\{\phi_i,\phi_j\}$. Furthermore there  
are several factorised boundary terms, which have the form 
\begin{eqnarray}
\Vev{\int\!\phi_i^{(2)}\al_{a_0}\al_{a_1}
 \int\!\al_{a_2}^{(1)}\cdots
 \int\!\al_{a_{k-1}}^{(1)}\int\!\al_{b}^{(1)}\int\!\al_{a_{l+1}}^{(1)}\cdots
 \int\!\al_{a_{n-1}}^{(1)}\al_{a_n}}\times&&
\nonumber\\
\times\Vev{\int\!\phi_j^{(2)}\al^{b}\al_{a_j}
 \int\!\al_{a_{j+1}}^{(1)}\cdots\int\!\al_{a_{l-1}}^{(1)}\al_{a_l}},&&
\end{eqnarray}
and similar terms with $i$ and $j$ interchanged, as 
depicted in \figref{bfact}. 
They can be written $\Phi_i\circ\Phi_j \pm \Phi_j\circ\Phi_i$ which 
can be understood as a supercommutator for higher order maps. 
This supercommutator therefore corresponds to the deforming operator 
$\{\phi_i,\phi_j\}$. 

Similarly, for the product we have to study the on-shell equality 
\begin{equation}
\Vev{\int\!(Q\phi_i)^{(2)}\int\!(Q\phi_j)^{(2)}\al_{a_0}\al_{a_1}
 \int\!\al_{a_2}^{(1)}\cdots\int\!\al_{a_{n-1}}^{(1)}\al_{a_n}} = 0.
\end{equation}
This one is a bit more involved because it reduces to a codimension 2 boundary. 
One boundary term now involves the product $\int\!(\phi_i\cdot\phi_j)^{(2)}$. 
The other boundary terms are factorisations, defining the product in terms 
of the maps $\Phi_i$ and $\Phi_j$. These boundary terms are of the form
\begin{equation}
\Vev{\al_{a_0}\al_b\al_c}
\Vev{\int\!\phi_i^{(2)}\al^{b}\al_{a_1}
 \int\!\al_{a_2}^{(1)}\cdots\int\!\al_{a_{k-2}}^{(1)}\al_{a_{k-1}}}
\Vev{\int\!\phi_j^{(2)}\al^{c}\al_{a_{k}}
 \int\!\al_{a_{k+1}}^{(1)}\cdots\int\!\al_{a_{n-1}}^{(1)}\al_{a_n}}.
\end{equation}
Therefore the map corresponding to the product $\phi_i\cdot\phi_j$ 
can formally be written in the form $m(\Phi_i,\Phi_j)$.

In conclusion, we found that we could connect to each closed string
operator a series of multilinear maps, which can be seen as the
deformations of the algebraic structure. Furthermore, we saw that the
algebraic structure of the \emph{deforming} closed string -- the $G$
algebra formed by $Q$, $m$ and $b$ -- is reflected by a corresponding
algebraic structure on the algebra of maps.

\section{Hochschild and Deformation Complexes}
\label{sec:defcplx}

In this section we study deformations of closed strings (2-algebras)
in a more abstract setting. We saw above that a (topological) closed
string theory has the structure of a Gerstenhaber algebra, formed by
the BRST operator $Q$, the OPE product $\cdot$ and the bracket
$\{\cdot,\cdot\}$. These are part of an algebra of multilinear maps;
this structure will play an essential role in the deformation theory
of the closed string algebra. Considering the deformation complex we
will find that there can be several different ways to deform this
algebra, depending on which part of the structure one wants to deform.

\subsection{The Hochschild Complex}
\label{hodef}

Mathematically, the deformation of an algebra $A$ is controlled by its
Hochschild complex $\Hoch(A)$. Let us first focus on associative
algebras $A$. Operations in $A$ are multilinear maps acting on the
vector space $A$. The vector spaces $C^n(A,A)=\Hom(A^{\otimes n},A)$,
consisting of $n$-linear maps in $A$, define the degree $n$ space of
what is known in mathematics as the Hochschild complex $\Hoch(A)$ of
the algebra $A$. Algebraic operations and differentials are special
elements in this space. Moreover, any deformation of the algebraic
structure is naturally an element of the Hochschild complex.

The Hochschild complex of an algebra has an interesting algebraic
structure by itself, which plays an important role in the deformation
theories. Part of this structure contains information about the
algebra $A$ that is deformed. We first extend the action of a map in 
$\Phi\in C^n(A,A)$ to the full tensor algebra $\CT A=\bigoplus_l A^{\otimes l}$. 
This action is defined as follows
\begin{equation}\label{action}
\Phi(\al_1, \ldots, \al_{l}) = \sum_{k=0}^{l-n} (-1)^{k(n-1)} 
 (\al_1,\ldots,\al_k,\Phi(\al_{k+1},\ldots,\al_{k+n}),\al_{k+n+1},\ldots,\al_{l}). 
\end{equation}
For graded algebras there are extra signs coming from $\Phi$ passing the $\al$'s. 
These are standard, and we will not include them in the notation. 
Through \eqref{action}, we reinterpret the maps in $C^n(A,A)$ as maps on $\CT A$, 
lowering the tensor degree by $n-1$.\footnote{We could 
have started by defining the maps in $C(A,A)$ by their action on the full 
tensor algebra. It can be shown however that a map on the tensor algebra 
lowering the degree by $n-1$ is completely determined by its lowest component, 
that is its action on $A^{\otimes n}$. Hence the definitions are equivalent.}
The composition of multilinear maps is thus a composition on the space $C^*(A,A)$; 
it is a fundamental operation on the Hochschild complex. 
The generating action of the composition of two elements $\Phi_i\in C^{n_i}(A,A)$, 
i.e.\ its action on $A^{\otimes (n_1+n_2-1)}$, is given by 
\begin{eqnarray}
&&(\Phi_1 \circ \Phi_2)(\al_1, \ldots, \al_{n_1 + n_2-1}) = \\
 &&\qquad\sum_{k=0}^{n_1-1} (-1)^{k(n_2-1)} 
\Phi_1 (\al_1,\ldots,\al_k,\Phi_2(\al_{k+1},\ldots,\al_{k+n_2}),\al_{k+n_2+1},\ldots,\al_{n_1+n_2-1}).
\nonumber
\end{eqnarray}
This definition makes the formulas in the previous section more precise. 

Using the composition as a product on $C(A,A)$, 
we can define a natural supercommutator called the bracket, 
which is defined by 
\begin{equation}\label{bracket}
[\Phi_1,\Phi_2] = \Phi_1\circ\Phi_2 -(-1)^{(n_1-1)(n_2-1)}\Phi_2\circ\Phi_1,
\qquad \Phi_i\in C^{n_i}(A,A).
\end{equation}
The order of the map minus one is interpreted as a degree. 
When $A$ is graded, the maps can also carry an extra grading from this, which 
would introduce standard extra signs in the definition above. 
It is easy to show that this bracket satisfies a graded version of the 
Jacobi identity, making the algebra of maps into a Lie algebra. Notice that 
the bracket lowers the total order of the maps by one. Because we interpret the 
order as a degree, the bracket has intrinsic degree $-1$, so that it is not a 
regular Lie bracket. 

Many familiar relations between algebraic operations can be 
rewritten elegantly in terms of this structure. The condition on a coboundary 
operator $Q$ is $Q\circ Q=0$, which can be written in terms of the bracket as $[Q,Q]=0$. 
The associativity of a bilinear product $m\in C^2(A,A)$ is equivalent to $[m,m]=2m\circ m=0$. 
The derivation condition of the product (Leibniz rule) can be written $[Q,m]=0$. 
If we consider a differential associative algebra, with product $m$ and differential $Q$, 
these three defining conditions (coboundary, derivation and associativity) can be written 
as the single equation $[Q+m,Q+m]=0$, by decomposing this into its separate degrees. 
Notice that although $Q+m$ does not make much sense as a multilinear map on $A$, 
it does as a map on  $\CT A$. This definition also 
makes it almost obvious to introduce $A_\infty$ algebras. An $A_\infty$ algebra is 
defined in terms of a set of multilinear products $m_n\in C^n(A,A)$, $n=1,2,\ldots$, 
such that $m_n$ has degree $2-n$. These maps should satisfy a generalised associativity 
condition, which in terms of the total sum $m=m_1+m_2+\cdots$ can be written 
$[m,m]=0$. By decomposing into the various degrees, this gives an infinite 
number of relations. For a differential associative algebra, $m_n=0$ for $n\geq 3$. 
In general, the first two conditions -- coboundary and Leibniz -- are not altered. 
However the associativity condition is changed by the trilinear product $m_3$ as follows 
\begin{equation}
m_2\circ m_2+m_1\circ m_3+m_3\circ m_1=0, 
\end{equation}
which is precisely the relation \eqref{assoc2} we found in the off-shell closed string. 
On the cohomology with respect to the differential $m_1$, the product $m_2$ reduces to 
an associative product. 

Let us assume that we deform a certain bilinear operator $m\in\Hom(A^{\otimes 2},A)$ 
(product or bracket), which has an internal degree $q$, satisfying a certain associativity 
or Jacobi constraint. Then we can build the following coboundary operator:
\begin{equation}\label{cob}
\delta_m\Phi = m\circ\Phi  +(-1)^{n+q|\Phi|} \Phi\circ m=[m,\Phi],
\qquad \Phi\in C^n(A,A). 
\end{equation} 
The coboundary condition $\delta_m^2=0$ follows from the associativity of the 
product $m$. Also, if we deform a linear operator $m_1\in\End(A)$ (which should be 
identified with $Q$) satisfying a coboundary constraint, 
it also acts on $C^*(A,A)$ as a coboundary operator. This coboundary 
operator, defined by $\delta_{m_1}=[m_1,\cdot]$, acts by conjugation on maps in $C(A,A)$. 
Actually, \eqref{cob} can be applied to any multilinear operation $m_n\in\Hom(A^{\otimes n},A)$, 
satisfying a generalisation of the associativity constraint, namely $m_n\circ m_n=0$. 
There may be several products that are deformed. The full 
complex $C^*(A,A)$ then is a multicomplex, having several coboundary operators. 
For a consistent deformation theory, all these coboundary operators need to 
commute. This will be guaranteed by additional constraint on the deformed 
operations (such as Leibniz). We can then construct a total coboundary operator 
$\delta$, which is a weighed sum of the several coboundary operators. 

Any product $m$ defines a cup product $\cup_m$ on the algebra 
$C^*(A,A)$ by the definition 
\begin{equation}\label{ghpr}
\cup_m(\Phi_1,\Phi_2)(\al_1, \dots , \al_{n_1+n_2}) = 
m\Bigl( \Phi_1(\al_1,\ldots,\al_{n_1}), \Phi_2(\al_{n_1+1}, \al_{n_1+n_2}) \Bigr),
\end{equation}
where $\Phi_i\in C^{n_i}(A,A)$. This can be generalised straightforwardly to 
general-order products $m_n\in C^n(A,A)$. The corresponding cup product $\cup_{m_n}$ 
is a product of order $n$, acting on the algebra $C(A,A)$. 

With the coboundary operator and the bracket, the Hochschild complex 
$C^*(A,A)$ is a (twisted) differential graded Lie algebra. 
Including the cup product makes it a differential Gerstenhaber algebra. 

We can repeat all of the above for the (graded) antisymmetric case, 
giving a generalisation of Lie algebras. The main difference is that we replace 
the tensor product $\CT A$ by the exterior algebra $\ext[*]A$. Maps on the 
exterior product therefore become antisymmetric multilinear  maps on $A$. 
The formula for the composition of maps then becomes a signed sum over all 
permutations of the arguments. For a single bilinear antisymmetric map $b$, 
this gives three terms in the formula for $b\circ b$. The vanishing 
of $b\circ b$ is equivalent to the Jacobi identity. 
Similarly, one can introduce a coboundary $Q$, and define a differential 
Lie algebra by the condition $[Q+b,Q+b]=0$. More generally, an $L_\infty$ 
algebra is defined by an infinite number of multilinear antisymmetric maps 
$b_n:\ext[n]A\to A$, $n=1,2,\ldots$,  of degree $2-n$ satisfying $[b,b]=0$, 
where $b=b_1+b_2+\cdots$. 

To see the relevance of this structure for deformations, we consider the deformation 
of an associative product $m$. We deform the product 
by an element $\Phi$. The fully deformed product is given by the correlation 
functions with an exponentiated insertion \eqref{deform}. The resulting deformed 
product is written $m+\Phi$. We want the deformed product to 
satisfy the generalised associativity condition, i.e. $(m+\Phi)^2=0$. Using the 
associativity of the undeformed product $m$, we find the condition 
\begin{equation}
\delta_m\Phi + \frac{1}{2}[\Phi,\Phi] = 0. 
\end{equation}
This formula is called the master equation, or Maurer-Cartan equation, 
of the deformation theory. To first order in the deformation parameter 
we find that $\Phi$ should be closed with respect to the coboundary operator 
$\delta_m$ on the Hochschild complex. We could make this more precise by making a 
Taylor expansion for the deformation, $\Phi = \sum_{n\geq1}t^n\Phi_{n}$, 
in terms of some deformation parameter $t$. This gives an infinite number 
of relations of the form 
\begin{equation}\label{recursion}
\delta_m\Phi_{n}= - \frac{1}{2}\sum_{n_1+n_2=n}[\Phi_{n_1},\Phi_{n_2}].
\end{equation}
Note that $n_1,n_2\geq1$; therefore this equation can be used to find
the higher order corrections to $\Phi$  recursively, as $\Phi_{n}$ does not 
occur on the right-hand side of \eqref{recursion}. For this, one needs to able to 
find a left inverse of the operation $\delta_m$, that is to solve the equation 
$\delta_m\Phi=\Psi$ for $\Phi$, with general $\Psi$. Any failure for this 
existence is an obstruction. It means that for a given 
infinitesimal deformation $\Phi_{1}$, one may not be able to find higher 
corrections in order to satisfy the full associativity. Or in other words, 
not every infinitesimal deformation may be extendable to a full deformation. 
This can only happen when the third cohomology with respect to $\delta_m$ is 
nonzero, since the right-hand side is in $C^3(A,A)$ and can be shown to be closed with 
respect to $\delta_m$. From this we see that the second cohomology of $\delta_m$ 
contains the infinitesimal deformations, and the third cohomology contains 
potential obstructions.

\subsection{Deformation Complexes of Closed Strings}

In the previous section we saw a way to deform the algebra of closed strings, 
by the insertion of extra integrated operators on the worldsheet. 
In this section we discuss deformations in the context of the 
deformation complex, which describes the basic cohomology theory 
governing the deformation of the algebra.
The deformation complex $\Def(A)$ of an algebra is a graded Lie 
algebra containing all possible deformations of this algebra. 
In any deformation theory of algebras, the central role in the 
deformation complex is played by the Hochschild complex, 
which we already met. The grading is such that $\Def^1(A)$ corresponds 
to the infinitesimal deformations of $A$, $\Def^0(A)$ contains the 
(global) symmetries, and $\Def^2(A)$ contains potential obstructions 
to extend the infinitesimal deformations to finite ones. The other 
gradings correspond to higher symmetries and higher obstructions. 
Generally, the deformation complex can be decomposed (as a vector space) 
as $\Def(A)=A\oplus C(A,A)$. The first factor $A$ is quite trivial, 
and corresponds to shifts of the elements of the algebra 
(translations in $A$). The second factor $C(A,A)$ is the 
Hochschild complex, containing deformations 
of the products. In the following we will ignore the first factor, 
as it will play no significant role in the discussion. The most 
important effect of this factor is that it kills the first factor 
$C^0(A,A)=A$ in the Hochschild cohomology, corresponding 
to maps of order 0.\footnote{As $C(A,A)$ corresponds to deformations 
of the background, and the first factor $A$ can be interpreted as a 
perturbation of the theory by the operators in the theory itself, 
we speculate that this cancellation should be interpreted physically
as background independence.}

Up to now, we have 
treated the algebra $A$ merely as a vector space, and we did not yet 
use any information about the product structure it may have. This 
information will supply the vector space $C(A,A)$ with some extra 
structure. Most important for the deformation theory is the fact that 
there will be a coboundary operator $\delta$ on $C(A,A)$, making it 
into a complex. This coboundary operator will precisely be determined 
by the algebraic structure that is deformed. Indeed, we saw earlier 
that the deformation of a bilinear product $m$ satisfying an associativity 
condition defines a coboundary operator $\delta_m$ on $C(A,A)$. Moreover, 
we saw that this coboundary operator was closely related to the 
deformation problem of the product $m$: the cohomology of appropriate degree 
describes the possible infinitesimal deformations. This shows 
precisely what we need in addition to define the deformation complex. 
We need a coboundary operator $\delta$ of degree 1 
with respect to an appropriate grading on the space $C(A,A)$. 
This coboundary operator is determined by the structure we are deforming. 
The grading also plays an important role. It determines how the 
different maps in the deformation complex should be interpreted; for example, 
the true deformations have degree 1, the elements of degree 0 are 
related to symmetries, and the degree 2 elements describe obstructions. 
Indeed, we know from examples in physics that the interpretation 
of several operations or operators can depend on the deformation problem 
one studies. 

If $A$ is merely a complex -- 
a graded vector space with a coboundary operator $Q$ of degree 1, 
it can be considered a 0-algebra (the state space of a point). 
In this case the only thing we can deform is $Q$, so we should take 
for the coboundary on the deformation complex the operator $\delta_Q$. 
The deformation complex now has the structure of a differential associative 
algebra. The product is given by the composition $\circ$, 
which obviously is associative. Actually, the only relevant part turns out 
to be $A\oplus\End(A)$, forming the algebra of affine transformations on 
$A$ \cite{kon2}. Hence the cohomology of the deformation complex of a 
0-algebra is an associative algebra, or a 1-algebra in the language of 
\cite{kon2}. 

If $A$ is an associative algebra, we deform the product $m$. Then the 
coboundary operator on the deformation complex is given by $\delta_m$. 
This deformation complex has the structure of a Gerstenhaber algebra, 
formed by the coboundary $\delta_m$, the cup product $\cup_m$, and the 
Gerstenhaber bracket. Hence the deformation complex of a 
1-algebra is a 2-algebra. If $A$ is a differential associative 
algebra, we can also deform its coboundary operator $Q$. This would supply 
the deformation complex with a second coboundary operator $\delta_Q$. 
The natural question then arises as to which one of the two 
coboundary operators defines the structure of complex for the 
deformation complex. The answer is both. To see how this works 
notice that  in this situation the vector space $C(A,A)$ has a 
double grading. One grading comes from the map degree, which we denote $n$, 
the other comes from the internal grading of $A$, let's call it $q$. 
The space of maps break up into doubly graded spaces $\Hom(A^{\otimes n},A)^q$. 
The two coboundary operators $\delta_Q$ and $\delta_m$ have bidegrees 
$(q,n)$ given by $(1,0)$ and $(0,1)$ respectively, and make $C(A,A)$ into 
a double complex. The essential condition that the two coboundary operators have 
to anticommute follows from the Leibniz rule. The total coboundary 
operator of the double complex is given by the sum. This also 
implies that the total degree on the complex is given by the sum, $p=q+n-1$, 
so that both coboundary operators raise the total degree by 1. Here the shift by 1 
is related to a mathematical convention, which requires degree $p=0$ 
in the deformation complex to correspond to symmetries, and we 
definitely want $\End(A)^0$ (reparametrisations) to be interpreted as 
symmetries of the algebra. Also, this fits nicely with the structure of the 
Hochschild complex, in which $n-1$ turns up as a natural grading. 

With all this in mind, we now turn to the deformation of a 2-algebra. 
We will be working mainly on-shell, therefore the algebraic structure 
is that of a differential Gerstenhaber algebra, as was considered in 
the beginning of \secref{sec:tcs}. As the BRST operator in general 
is deformed we also need to include it in our discussion. We know that 
off-shell structure should be a homotopy algebra, but we will assume  
that we can work in this restricted setting.\footnote{Using operad 
descriptions of the more general off-shell structure one can in principle 
define a more general deformation theory for these. However this becomes much 
more involved, and will be far beyond the scope of this paper.}
In this situation, there are three operators which we can potentially 
deform. This gives us three different coboundary operators, 
$\delta_{Q}$, $\delta_m{}$, $\delta_b$, on the space $C(A,A)$. 
Corresponding to these coboundaries, there are three types of arrows 
in the complex. The diagonal ones  come from the product, the vertical 
ones from the BRST operator, and the horizontal ones from the bracket. 
The deformation complex therefore looks as follows. 
\begin{equation}
\matrix{
A^{0}             & \labrarr{\delta_b} & \hoch[-1]{1}& \rarr    & \hoch[-2]{2}& \rarr    & \cdots\cr
\labdarr{\delta_Q}& \labsearr{\delta_m}& \darr       & \searrow & \darr       & \searrow &       \cr
A^{1}             & \rarr              & \hoch[0]{1} & \rarr    & \hoch[-1]{2}& \rarr    & \cdots\cr
\darr             & \searrow           & \darr       & \searrow & \darr       & \searrow &       \cr
A^{2}             & \rarr              & \hoch[1]{1} & \rarr    & \hoch[0]{2} & \rarr    & \cdots\cr
\darr             & \searrow           & \darr       & \searrow & \darr       & \searrow &       \cr
A^{3}             & \rarr              & \hoch[2]{1} & \rarr    & \hoch[1]{2} & \rarr    & \cdots\cr
\darr             & \searrow           & \darr       & \searrow & \darr       & \searrow &        
}
\end{equation}
The natural thing to do now is to define 
a total coboundary which is basically the sum of the three. However, 
it is impossible to define a degree on the complex such that all three 
maps have degree 1. Physically, this means that we cannot for example 
identify gauge symmetries and true deformations for all three operators 
simultaneously. This implies that we cannot consistently deform 
all three operators at the same time. What can happen physically is that 
deformations of one operator (corresponding to deformations of degree 
1 in the deformation complex) are obstructions for 
deformations of another operator (having degree larger than 1 in 
that deformation complex), or otherwise (gauge) symmetries 
(having degree smaller than 1). 

It is possible to deform two structures at the same 
time. Then we keep two of the three arrows in the complex, and we find a 
double complex. First note, that we can introduce a pair of quantum 
numbers, $(p,q)$ say, such that one of the maps has quantum numbers 
$(1,0)$, and the other $(0,1)$. For example, if we keep the vertical 
and horizontal arrows, we can take the row and column numbers. The 
total degree then is simply the sum of the two, so that both 
maps indeed have degree 1. The total differential is more or less 
the sum of the maps (up to some relative signs). The degrees should 
always be chosen such that the space $\End(A)^0$ has degree $(0,0)$, 
as indeed these should certainly be interpreted as gauge symmetries. 
From the point of view of the algebra, we are really deforming only a 
substructure. The three substructures we can deform correspond to 
the differential associative (DA) structure, the differential Lie 
(DL) structure, and the Gerstenhaber (G) structure.

\section{Classification of Closed String Deformations}
\label{sec:cplxes}

In this section we will discuss the three possibilities for deforming
the structure of a closed string algebra separately.

The deformation complex breaks down into the vector spaces 
$\Hom(A^{\otimes n},A)^q$, where $n$ is the order and $q$ denotes the 
internal ghost degree (that is, a corresponding map raises the internal degree 
in $A$ by $q$). The operations $m$ we want to deform are particular elements 
in this space, so they also carry the corresponding degrees. It is easily seen 
that if $m$ has ghost degree $q$ and order $n$, than the corresponding 
coboundary operator $\delta_m$ increases the ghost degree by $q$ 
and the order by $n-1$. 
From this now we can derive the expression of the total degree in the deformation complex. 
The necessary condition is that for each operator $m$ that is deformed, the corresponding 
coboundary operator $\delta_m$ should have total degree $1$.
Here we use the degrees $(q,n)$ of the various operations: $\delta_Q$ has degrees 
$(1,0)$, $\delta_m$ has degree $(0,1)$, and $\delta_b$ has degrees 
$(-1,1)$. The various possibilities for choosing the degree in such a way that two 
operations are deformed are given in \tbref{tb:defs}. The offset of the degree 
is determined by the fact that obviously $\End(A)^0\subset\Def^0(A)$. 
\begin{table}
\begin{center}
\begin{tabular}{|c|c|c|}
\hline
Algebra & Total coboundary & Total degree $p$ \\
\hline
DA & $\delta_m+\delta_Q$ & $q+n-1$ \\
DL & $\delta_b+\delta_Q$ & $q+2n-2$ \\
G  & $\delta_b+\delta_m$ & $n-1$ \\
\hline
\end{tabular}
\end{center}
\caption{The three deformations of a $2$-algebra: differential algebra, 
differential Lie algebra and Gerstenhaber algebra. The last column gives 
the formula for the total degree $p$ in the deformation complex, in terms 
of the internal ghost number $q$ and the order $n$.}
\label{tb:defs}
\end{table}
Before we turn to a description of the three cases, we will first
examine the significance of the degrees.

\subsection{Gradings and Dimensions}

 In general the gradings have the form $p=\al q+\beta(n-1)$, 
the most general linear relation between the degrees such that $\End(A)^0$ 
corresponds to degree $p=0$. We have to be careful with the definition 
of these degrees. In general, the degrees $p$ and $q$ refer to the 
ghost numbers of the zeroth descendants, modulo a shift in the definition 
of $p$. For example, for the bracket $\{\al_a,\al_b\}$ the degree 
is given  by $q=g_{\{\al_a,\al_b\}}-g_{\al_a}-g_{\al_b}=-1$. 

We want to argue here that the coefficient 
$\beta$ is related to the dimensionality of the deforming theory. 
To see this, let us look at more general topological field theories 
in any dimension $d$. They always come with a Lie bracket, which is the 
generalisation of the bracket in two dimensions, and is defined by 
\begin{equation}
\{\phi_1,\phi_2\} = \oint_{C}\phi_1^{(d-1)}\phi_2,
\end{equation}
where $C$ is a $(d-1)$-cycle enclosing the insertion point of $\phi_2$ 
(a $(d-1)$-sphere). 
For $d=1$ this gives the commutator with respect to the product, 
as the cycle $C$ consists of the (formal) difference of two points. 
This Lie bracket has degree $1-d$, due to the descendant. Restricted to 
the BRST-closed operators, it is easily seen to be independent 
of the choice of the cycle $C$ and to satisfy the Jacobi identity. 

There is a natural relation between this Lie bracket and the 
quantum commutator in canonical quantisation. In a canonical quantisation, 
we use a time slicing for our space-time, and a time-coordinate $x^0$. 
Assume two canonically quantised operators $\hat\phi_1$ and $\hat\phi_2$
satisfying a commutation relation of the form 
\begin{equation}
[\hat\phi_1^{(d-1)}(y),\hat\phi_2(x)] = \hat\phi_3(x)\delta(y-x).
\end{equation}
Here the $(d-1)$th descendant is natural, because the delta function 
should be considered a $(d-1)$-form. If we want to calculate the 
Lie bracket defined above in a canonical quantisation, we 
should split the cycle $C$ up into two half-spheres $C=D_+\cup D_-$, 
at times $y^0>x^0$ and $y^0<x^0$, according to the time-slicing we chose. 
We can deform these half-spheres to two space-slices, pushing the 
strip on the side to infinity, where it should not give any contribution. 
The quantisation of the bracket can then be written 
\begin{equation}
\{\hat\phi_1,\hat\phi_2\}(x) 
= \int_{D_+}\hat\phi_1^{(d-1)}(y)\hat\phi_2(x) 
  - \int_{D_-}\hat\phi_1^{(d-1)}(y)\hat\phi_2(x) 
=\int_D\hat\phi_3(x)\delta(y-x) = \hat\phi_3(x), 
\end{equation}
where we used the quantum commutator above. Hence we see that indeed 
the quantum commutator directly maps to the Lie bracket. This 
procedure is well-known in the context of two-dimensional 
CFT's, where it describes the action of currents. In a canonical 
quantisation we can therefore relate the operator $\phi_1$ 
to a differential operator $\phi_3\frac{d}{d\phi_2}$. Because of the 
descendant in the definition of the bracket the operator $\phi_1$ 
and the corresponding map differ in degree by an amount of $d-1$. 
This indeed corresponds to $\beta=d-1$. 
 
If we consider a pair of canonically conjugate operators $\phi_2=\phi$ 
and $\phi_1=\pi$, we have $\phi_3=1$. Let us for simplicity work 
in a first-order formalism, where $\pi$ and $\phi$ are both fundamental 
fields. In this case it is straightforward 
to connect to the Hochschild complex. The canonical quantisation 
gives $\pi^{(d-1)}\sim\frac{d}{d\phi}$, which shows that the operator 
$\pi$ corresponds to an element of $\Hom(A,A)$ in the Hochschild 
complex. More generally, the operator $\pi^n$ gives an element 
in $\Hom(A^{\otimes n},A)$. Let us compare the various degrees.
The degrees refer explicitly to the degrees of the zeroth descendants 
of the operators, except for the degree $p$ in the deformation complex, 
which in $d$ dimensions is shifted by $d-1$. Denote the ghost degrees of 
$\phi$ and $\pi$ as $g_\phi$ and $g_\pi$ respectively. In the 
action there should be a term of the form $\int\pi^{(d-1)}d\phi$ 
as we are working in a first-order formalism. This implies that 
$g_\pi=-g_\phi+d-1$. An operator of the form $\pi^n$ (and its descendants) 
now corresponds to an element in the deformation complex of 
degree $p=n g_\pi-d+1$. The induced multilinear map has a component 
in the maps of degree $n$ which acts as $\Bigl(\frac{d}{d\phi}\Bigr)^{\otimes n}$,  
which has an explicit ghost number $q=-n g_\phi$. Comparing the two 
degrees we find 
\begin{equation}\label{degform}
p = ng_\pi-d+1=n(-g_\phi+d-1)-d+1=q+(d-1)(n-1). 
\end{equation}
More generally, the operator $\pi^{n}$ induces other multilinear maps 
of the form $\pi^m\Bigl(\frac{d}{d\phi}\Bigr)^{\otimes (n-m)}$, which 
are in $\Hom(A^{\otimes(n-m)},A)^q$, where $q=mg_\pi-ng_\phi$. These 
maps can be considered as different descendants of the operator $\pi^n$. 
Comparing the degrees one finds 
exactly the same relation, if we replace $n$ by the order $n-m$ of the map. 
We conclude that the coefficient $\beta$ equals $d-1$. 

Let us comment on the shift in degree in the deformation complex. 
This has to do with the mathematical convention for the degree in the 
deformation complex. This is such that the actual deformations have 
degree 1. These operators should however correspond to the physical 
operators, which in $d$ dimensions have ghost number $d$. This is 
because the corresponding perturbation of the action, $\int\phi^{(d)}$, 
should have ghost number 0. Therefore, we have to shift the degree by $d-1$, 
$p=g_\phi-d+1$. The mathematical degree can 
be considered the degree of the $(d-1)$th descendant of the operator, 
which defines a perturbation of a ``pre-Lagrangian'' $\widetilde L$, 
which is defined such that $S=\dint{dt}\widetilde L^{(1)}$.

\subsection{Deformation of the Differential Associative Structure}

Of the three possibilities, we first consider the deformation of the DA structure 
(differential algebra, or more generally the $A_\infty$ structure). 
The DA structure is determined by the symmetric 
product and the BRST operator $Q$. 
In physics this is the best-known problem, and in the context of
topological strings it gives rise to the WDVV equations 
\cite{dvv}. It is basically the problem of deforming the closed string 
using closed string operators. There is also a deformation of 
the BRST operator $Q$, which was studied in this context already 
\cite{verl}. The two structures together define the bracket in the 
usual way. But as is known, deformations of the closed string by closed 
string operators do not deform the bracket. Therefore, we expect that 
in general the bracket will be fixed and is not deformed. 

The deformation double complex for the deformation of the DA structure 
has the following structure 
\begin{equation}
\matrix{
\darr             &                   & \darr      &       & \darr      &       &       \cr
A^{0}             & \labrarr{\delta_m}& \hoch[0]{1}& \rarr & \hoch[0]{2}& \rarr & \cdots\cr
\labdarr{\delta_Q}&                   & \darr      &       & \darr      &       &       \cr
A^{1}             & \rarr             & \hoch[1]{1}& \rarr & \hoch[1]{2}& \rarr & \cdots\cr
\darr             &                   & \darr      &       & \darr      &       &       \cr
A^{2}             & \rarr             & \hoch[2]{1}& \rarr & \hoch[2]{2}& \rarr & \cdots\cr
\darr             &                   & \darr      &       & \darr      &       &
}
\end{equation}
The vertical arrows correspond to the coboundary $\delta_Q$, while the 
horizontal arrows correspond to $\delta_m$. 
So we see that the two gradings have a very natural interpretation: 
one (related to the BRST operator) is the internal ghost degree (target space 
degree), and the other (related to the product) is the map degree of the 
multilinear maps (the number of elements in the algebra on which it acts). 

The degree in the deformation complex is given by $p=n-1+q$. This 
means that the degree $p$ cocycles in the deformation complex are given by 
the elements of the space 
\begin{equation}
\Def^p(A) = \bigoplus_{n\geq0}\Hom(A^{\otimes n},A)^{p-n+1}.
\end{equation}
The most important part of the deformation complex is the degree 1 space. 
These contain the actual deformations of the algebra. This space is given by 
\begin{equation}
\Def^1(A) = A^{2}\oplus\End(A)^{1}\oplus\Hom(A^{\otimes 2},A)^0\oplus\Hom(A^{\otimes 3},A)^{-1}\oplus\cdots.
\end{equation}
This is very natural from the point of view of the string algebra. 
The terms in the physical deformations contain the deformed operations. 
For example the deformation of the BRST operator is an element of $\End(A)^1$, and the 
deformation of the product is an element of $\Hom(A^{\otimes 2},A)^0$. 

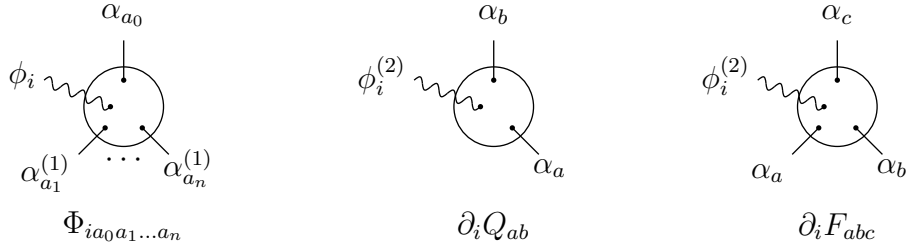
\begin{figure}
\begin{center}
\begin{picture}(345,100)
\put(0,0){
  \begin{picture}(85,100)(-5,-10)
    \CArc(40,45)(15,0,360)
    \Photon(10,55)(35,45)24  \Vertex(35,45)1  \Text(7,57)[r]{$\phi_{i}$}
    \Line(23,27)(33,37)  \Vertex(33,37)1  \Text(20,20)[r]{$\al_{a_1}^{(1)}$}
    \Line(57,27)(47,37)  \Vertex(47,37)1  \Text(55,22)[l]{$\al_{a_n}^{(1)}$}
     \Text(40,25)[m]{$\cdots$}
    \Line(40,55)(40,70)  \Vertex(40,55)1  \Text(40,80)[m]{$\al_{a_0}$}
    \Text(40,0)[m]{$\Phi_{ia_0a_1\ldots a_n}$}
  \end{picture}
}
\put(130,0){
  \begin{picture}(85,100)(-15,-10)
    \CArc(40,45)(15,0,360)
    \Photon(10,55)(35,45)24  \Vertex(35,45)1  \Text(7,57)[r]{$\phi_{i}^{(2)}$}
    \Line(57,27)(47,37)  \Vertex(47,37)1  \Text(55,22)[l]{$\al_{a}$}
    \Line(40,55)(40,70)  \Vertex(40,55)1  \Text(40,80)[m]{$\al_{b}$}
    \Text(40,0)[m]{$\del_i Q_{ab}$}
  \end{picture}
}
\put(260,0){
  \begin{picture}(85,100)(-15,-10)
    \CArc(40,45)(15,0,360)
    \Photon(10,55)(35,45)24  \Vertex(35,45)1  \Text(7,57)[r]{$\phi_{i}^{(2)}$}
    \Line(23,27)(33,37)  \Vertex(33,37)1  \Text(20,20)[r]{$\al_{a}$}
    \Line(57,27)(47,37)  \Vertex(47,37)1  \Text(55,22)[l]{$\al_{b}$}
    \Line(40,55)(40,70)  \Vertex(40,55)1  \Text(40,80)[m]{$\al_{c}$}
    \Text(40,0)[m]{$\del_i F_{abc}$}
  \end{picture}
}
\end{picture}
\end{center}
\caption{A typical diagram corresponding to an element of the 
Hochschild cohomology for a deformation of the associative structure, 
the three-point function giving the deformation of the BRST operator, 
and the four-point function giving the deformation of the product.}
\label{defDA}
\end{figure}

The formula for the total degree \eqref{degform} suggests that we should 
take the first descendants for the $n$ incoming closed string operators in 
a string diagram corresponding to $\Phi_{ia_0\ldots a_n}$. 
This is indicated in \figref{defDA}. 
This is precisely the structure we found in the case of WDVV. 
Also, the formula for the degrees matches up exactly with the one 
for WDVV, \eqref{ghostWDVV}, if we take into account the 
remarks of the last subsection: the definition of the degree 
of the map matches exactly, $q=g_\Phi$, while the ghost degree of 
$\phi$ is shifted by $1=d-1$, so that $p=g_\phi-1$. 

The Hochschild cohomology has the structure of a Gerstenhaber 
algebra, with the bracket having degree $-1$. We saw that for 
deformation of the closed string by itself this algebra could 
be identified with the (on-shell) algebra of the closed string 
itself. We conclude that the WDVV equations describe a deformation 
theory of the DA structure.

\subsection{Deformation of the Differential Lie Structure}

Secondly, we consider the deformation of the differential Lie algebra 
structure, formed by the BRST operator $Q$ and the bracket. Now we find 
for the deformation complex the following form. 
\begin{equation}
\matrix{
\darr             &                   & \darr       &       & \darr       &       &       \cr
A^{0}             & \labrarr{\delta_b}& \hoch[-1]{1}& \rarr & \hoch[-2]{2}& \rarr & \cdots\cr
\labdarr{\delta_Q}&                   & \darr       &       & \darr       &       &       \cr
A^{1}             & \rarr             & \hoch[0]{1} & \rarr & \hoch[-1]{2}& \rarr & \cdots\cr
\darr             &                   & \darr       &       & \darr       &       &       \cr
A^{2}             & \rarr             & \hoch[1]{1} & \rarr & \hoch[0]{2} & \rarr & \cdots\cr
\darr             &                   & \darr       &       & \darr       &       &       \cr
A^{3}             & \rarr             & \hoch[2]{1} & \rarr & \hoch[1]{2} & \rarr & \cdots\cr
\darr             &                   & \darr       &       & \darr       &       &       
}
\end{equation}
The vertical arrows are again determined by $\delta_Q$; the horizontal 
arrows correspond to $\delta_b$ in this case. 
The total degree in the deformation complex is given by $p=2n-2-q$.
The actual deformations, that is the deformations of degree one, 
are given by 
\begin{equation}
\Def^1(A) = A^{3}\oplus\End(A)^{1}\oplus\Hom(A^{\otimes 2},A)^{-1}
  \oplus\Hom(A^{\otimes 3},A)^{-3}\oplus\cdots.
\end{equation}
The terms in the deformation complex at degree one show which operations 
are potentially deformed. We find the maps of degree one, corresponding 
to deformations of the BRST operator, and bilinear maps of degree $-1$, 
indicating the deformation of the bracket. The next term, that is 
trilinear maps of degree $-3$, will also play an important role in the 
deformation theory, as we will see below. This is the deformation 
complex that is most natural from the mathematical point of view, and 
in the mathematics literature it is referred to as describing 
the deformations of a Gerstenhaber algebra \cite{kon2,tam}. 
The Hochschild cohomology has the structure of a Poisson algebra, 
as also found in \cite{kon2}. 
The degree of the Poisson bracket, which is given by \eqref{bracket}, 
has degree $-2$, which is even. 

\begin{figure}
\begin{center}
\begin{picture}(345,100)
\put(0,0){
  \begin{picture}(85,100)(-5,-10)
    \CArc(40,45)(15,0,360)
    \Photon(10,55)(35,45)24  \Vertex(35,45)1  \Text(7,57)[r]{$\phi_{i}$}
    \Line(23,27)(30,34)  \Vertex(30,34)1  \Text(20,20)[r]{$\al_{a_1}^{(2)}$}
    \Line(57,27)(50,34)  \Vertex(50,34)1  \Text(55,22)[l]{$\al_{a_n}^{(2)}$}
     \Text(40,25)[m]{$\cdots$}
    \Line(40,60)(40,70)  \Vertex(40,60)1  \Text(40,80)[m]{$\al_{a_0}$}
    \Text(40,0)[m]{$\Phi_{ia_0a_1\ldots a_n}$}
  \end{picture}
}
\put(130,0){
  \begin{picture}(85,100)(-15,-10)
    \CArc(40,45)(15,0,360)
    \Photon(10,55)(35,45)24  \Vertex(35,45)1  \Text(7,57)[r]{$\phi_{i}^{(2)}$}
    \Line(57,27)(50,34)  \Vertex(50,33)1  \Text(55,22)[l]{$\al_{a}$}
    \Line(40,60)(40,70)  \Vertex(40,60)1  \Text(40,80)[m]{$\al_{b}$}
    \Text(40,0)[m]{$\del_i Q_{ab}$}
  \end{picture}
}
\put(260,0){
  \begin{picture}(85,100)(-15,-10)
    \CArc(40,45)(15,0,360)
    \Photon(10,55)(35,45)24 \Vertex(35,45)1  \Text(7,57)[r]{$\phi_{i}^{(3)}$}
    \Line(23,27)(30,34)  \Vertex(30,34)1  \Text(20,20)[r]{$\al_{a}^{(1)}$}
    \Line(57,27)(50,34)  \Vertex(50,34)1  \Text(55,22)[l]{$\al_{b}$}
    \Line(40,60)(40,70)  \Vertex(40,60)1  \Text(40,80)[m]{$\al_{c}$}
    \Text(40,0)[m]{$\del_i G_{abc}$}
  \end{picture}
}
\end{picture}
\end{center}
\caption{A typical diagram corresponding to an element of the 
Hochschild cohomology for a deformation of the differential Lie structure, 
the three-point function giving the deformation of the BRST operator, 
and the four-point function giving the deformation of the bracket.}
\label{defDL}
\end{figure}
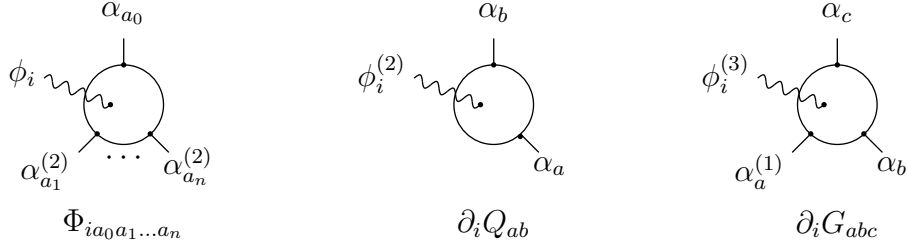

Following \eqref{degform}, we expect that this deformation theory 
should be considered as a 3-dimensional theory. The shift 
by $2n-2$ is typical for a 3-dimensional theory. We will argue in 
the next section that indeed this deformation theory enters naturally 
in the topological open membrane. The shift by $2n$ also indicates 
that for the higher correlation functions corresponding to the 
deformation complex, the extra insertions come as first descendants, 
as suggested in \figref{defDL}. 

\subsection{Deformation of the Gerstenhaber Structure}

Lastly, we consider the deformation of the Gerstenhaber structure,  
consisting of the product and the bracket. 
The deformation double complex now has the following form 
\begin{equation}
\matrix{
           &       &            &       &                   &                   & \darr        &       &       \cr
           &       &            &       & \hoch[0]{0}       & \labrarr{\delta_b}& \hoch[-1]{1} & \rarr & \cdots\cr
           &       &            &       & \labdarr{\delta_m}&                   & \darr        &       &       \cr
           &       & \hoch[1]{0}& \rarr & \hoch[0]{1}       & \rarr             & \hoch[-1]{2} & \rarr & \cdots\cr
           &       & \darr      &       & \darr             &                   & \darr        &       &       \cr
\hoch[2]{0}& \rarr & \hoch[1]{1}& \rarr & \hoch[0]{2}       & \rarr             & \hoch[-1]{3} & \rarr & \cdots\cr
\darr      &       & \darr      &       & \darr             &                   & \darr        &       &       \cr
\hoch[2]{1}& \rarr & \hoch[1]{2}& \rarr & \hoch[0]{3}       & \rarr             & \hoch[-1]{4} & \rarr & \cdots\cr
\darr      &       & \darr      &       & \darr             &                   & \darr        &       &     
}
\end{equation}
Now the arrows correspond to $\delta_m$ and $\delta_b$ respectively. 
The total degree is the degree of the map (the number 
of elements on which it acts), modulo a shift. The internal 
degree in the algebra $A$ does not contribute to the degree of the 
deformation complex. 

The degree in the deformation complex is given by $p=n-1$.
The gauge symmetries and matter content are given by the zeroth and 
first degree deformations, respectively:
\begin{equation}
\Def^0(A) = \End(A), \qquad
\Def^1(A) = \Hom(A^{\otimes 2},A).
\end{equation}
The grading is the same as for the Hochschild complex, apart 
from a shift by one. This implies that the bracket has degree 
$-2$, and the Hochschild cohomology has the structure of 
a Poisson algebra. The ghost degree does not play any role 
in the deformation theory. Therefore, we expect that the original 
ghost number symmetry is broken in this case.

\section{Topological Open Membranes and Boundary Strings}
\label{sec:bdystr}

In \cite{homaOS}, we studied deformations of boundary theories for open 
strings by bulk operators. We found that the deformation theory of this 
1-algebra indeed had the structure of a 2-algebra. This would lead us to 
expect that the 3-algebra deformation of the 2-algebra formed by the closed 
strings can be found in the context of open membranes. In this section we will 
argue that this is indeed the case.

To see this we interpret the closed strings as the boundary theory of a topological 
open membrane, or TOM for short. The relevant algebra will be the algebra of 
boundary operators, and has the structure of a 2-algebra. Indeed, this is a closed 
string theory. The BRST operator $Q$ of the open membrane descends to this boundary 
string, by integrating the corresponding current over a half-sphere enclosing the 
boundary operator. The deforming algebra, formed by the $\phi_i$, is the bulk 
algebra of the membrane.

\subsection{Three-Dimensional Topological Field Theories}

Three-dimensional topological field theories can be treated in a manner quite 
similar to two-dimensional ones, so we will be quite brief here. There are 
three-point functions  $C_{ijk}$ defining a symmetric product, which are 
equivalent to the two-dimensional ones. Using a unit operator we define a 
metric by the two-point function equivalent to $C_{0ij}$. The bracket is now 
defined by the three-point functions 
\begin{equation}
B_{ijk} = \Vev{\phi_i\oint\!\phi_j^{(2)}\phi_k},
\end{equation}
where we integrate over a 2-sphere enclosing $\phi_k$. 
 
As for any TFT, we demand the presence of a BRST operator $Q$ and of an
operator $G$, such that $\{Q,G\}=d$. In the presence of a boundary, 
these operators also induce an action on the boundary operators, 
though in general there may be extra boundary terms. 
The symmetry current $G$ in the topological open membrane 
induces a Ward identity of the form
\begin{eqnarray}\label{ward}
0 &=& \sum_m\xi^\mu(x_m)\Vev{ \prod_n\phi_{i_n}(z_n)
   \al_{a_1}(x_1)\cdots
   G_\mu\al_{a_m}(x_m)\cdots\al_{a_r}(x_r)} \nonumber\\
 &&+ \sum_n\xi^\mu(z_n)\Vev{ \phi_{i_1}(z_1)\cdots G_\mu\phi_{i_n}(z_n)
   \cdots\phi_{i_s}(z_s)\prod_m\al_{a_m}(x_m)},
\end{eqnarray}
where the $z$'s are points in the bulk and the $x$'s are points on the boundary. 
Here the operators $\phi$ and $\al$ can be any operator, not necessarily 
BRST-closed. They can also be descendants. 
In this equation, $\xi^\mu$ is a globally defined conformal vector field. 
The conformal group of the 3-ball is $\SO(2,2)$, which is six-dimensional. 
Therefore, we have a basis of 6 vector fields to choose for the $\xi$'s. 
This counting relies very much on a conformally invariant gauge fixing of 
the open membrane. A priori we do not know if such a gauge fixing does exist. 
In the following we will assume this.

\subsection{Deformations}

We study deformations of the closed string correlation 
functions by including new operators $\phi_i$ in the correlation functions, 
which we view as deforming operators. However, these operators will now be 
bulk operators for the membrane. We can define mixed two-point functions by 
\begin{equation}
\Phi_{ia} = \Vev{\phi_i\al_a}.
\end{equation}
The mixed three-point functions are defined by  
\begin{equation}
\Phi_{iab} = \Vev{ \phi_i\al_{a}\int\!\al_b^{(2)}}.
\end{equation}
Notice that we cannot have any correlators ``in between''; if we would insert 
a first descendant, integrated over a cycle, we could always shrink the cycle to zero. 
Higher mixed correlators are given by 
\begin{equation}\label{higher1tom}
\Phi_{ia_0a_1\ldots a_n} =  
\Vev{ \phi_i\al_{a_0}
      \int\!\al_{a_1}^{(2)}\cdots\int\!\al_{a_n}^{(2)} }.
\end{equation}
We will assume that the closed string Ward identities for $G$ are still valid,
so that these correlators are symmetric in the closed string indices. For the 
relevant situations, we will argue below that this is indeed the case. 

When we introduce extra membrane operators in the $\Phi$'s, we should integrate them, 
\begin{equation}\label{higher2tom}
\Phi_{ija_0a_1\ldots a_n} =  
\Vev{ \phi_i\int\!\phi_j^{(3)}\al_{a_0}
      \int\!\al_{a_1}^{(2)}\cdots\int\!\al_{a_n}^{(2)} }.
\end{equation}
Now the algebra of deforming operators is assumed to have the same structure 
as the closed string theory. That is, we have $Q$ and $G$. Also, these operators 
should be related to the corresponding operators on the closed string theory. 
This would mean that the correlators are also symmetric in the $i,j$ indices. 
This should also be true if we introduce extra integrated deforming operators. 
Indeed, the $G$ operator is zero on these top forms. These assumptions imply that 
the mixed correlators are integrable: there are functions $\Phi_{a_0\ldots a_n}(t)$ 
such that $\Phi_{ia_0\ldots a_n}(t) = \del_i \Phi_{a_0\ldots a_n}(t)$, where 
$\del_i=\frac{\del}{\del t^i}$. The coefficients in the expansion in $t$ 
are the higher correlation functions. We can therefore formally write these 
deformed correlators as
\begin{equation}\label{deformtom}
\Phi_{a_0a_1\ldots a_n}(t) = 
\Vev{ \al_{a_0}\int\!\al_{a_1}^{(1)}\al_{a_2}
  \int\!\al_{a_3}^{(2)}\cdots\int\!\al_{a_n}^{(2)} 
  \,\e^{\,t^i\!\!\int\phi_i^{(3)}}}.
\end{equation}

\subsection{The Algebraic Structure of Open Membranes}

The essential identity needed to view the insertions of bulk operators 
as a deformation of the boundary algebra was the symmetry of the higher 
correlators $\Phi_{ija_0a_1\ldots a_n}$ defined in \eqref{higher2tom}, with 
respect to the bulk indices:
\begin{equation}
\Vev{\phi_i\phi_j^{(3)}\al_{a}} = C \Vev{\phi_i^{(3)}\phi_j\al_a}.
\end{equation}
where $C$ should be a function of the insertion points. In order for the 
integrated correlation functions to be truly invariant under this switch, 
this function should be the Jacobian of the coordinate transformation 
from the insertion point of $\phi_i$ to the insertion point of 
$\phi_j$.\footnote{Conformal invariance guarantees the existence of this 
coordinate transformation.} We will now argue that the assumption of conformal 
invariance gives enough global Ward identities to give the above relation at least. 
We are however not in a position to determine the factor $C$, due to a lack of 
understanding of the conformal invariance. Therefore the invariance of the 
integrated correlation functions will not be established completely. 
As we argued, assuming conformal invariance we have 6 independent Ward 
identities of the form \eqref{ward}. However, in the present case we 
do not want the boundary operator $\al_a$ to get involved. This 
can be established if the vector field $\xi^\mu$ used in the Ward identity 
is 0 at the insertion point of this operator. This gives two restrictions 
on $\xi$, leaving us with 4 Ward identities. These are however sufficient 
to transfer the 3 independent components of $G_\mu$ from $\phi_j$ 
to $\phi_i$, thereby establishing the existence of the above relation. 
As $G$ is 0 on any second descendant of a boundary operator or a 
third descendant of a bulk operator, the relation remains true if we 
insert any number of these maximal descendants. 

More important is a relation of the form 
\begin{equation}\label{ward2}
\Bla \int\!\phi_i^{(3)}\al_a  \oint\!\al_b^{(1)} \al_c \Bra 
= \Bla \phi_i \int\!\al_a^{(2)} \int\!\al_b^{(2)} \al_c \Bra,
\end{equation}
showing that we can interpret the mixed correlation functions 
as bulk to boundary metrics deformed by the boundary operators. 
This can be proved using Ward identities of the form
\begin{equation}
\Bla \phi_i^{(3)} \al_a \al_b^{(1)} \al_c \Bra 
= C \Bla \phi_i \al_a^{(2)} \al_b^{(2)} \al_c \Bra.
\end{equation}
We start from the right-hand side. A priori, 
we have 6 independent global vector fields. 
Next we choose the vector fields that fix the position of $\al_c$. 
As this gives two conditions, there are 4 vector fields. 
Of these, we use two vector fields to transfer the second descendant 
from $\al_b$ to $\phi_i$. Next we choose the third vector field such 
that it fixes the position of $\al_b$ as well (as this gives two more 
conditions, there are two independent choices). We can use this to
transfer one descendant of $\al_a$ to $\phi_i$ without getting 
additional terms. This argument shows that conformal invariance of the TOM 
theory is large enough to get this Ward identity (in fact, we only need 5 
independent vector fields). Again, we cannot decide whether $C$ is a Jacobian. 
We expect this to be true on the general basis of conformal invariance and will 
assume it henceforth. Equation \eqref{ward2} means that the correlator 
$\Phi_{iabc}$ is a deformation of the bracket. It would remain valid when 
we include extra fully integrated bulk and boundary insertions.

\begin{figure}
\begin{center}
\begin{picture}(235,90)
\put(0,5){%
  \begin{picture}(85,80)(-15,10)
    \CArc(40,45)(15,0,360)
    \Photon(10,55)(35,45)24  \Vertex(35,45)1  \Text(8,55)[r]{$Q\phi_i$}
    \Line(23,27)(30,34)  \Vertex(30,34)1  \Text(25,22)[r]{${a_1}$}
    \Line(57,27)(50,34)  \Vertex(50,34)1  \Text(55,22)[l]{${a_n}$}
    \Line(40,60)(40,70)  \Vertex(40,60)1  \Text(40,78)[m]{${a_0}$}
    \Text(40,25)[m]{$\cdots$}
  \end{picture}
}
\put(95,40){$\displaystyle =\,\sum_{k,l}\pm$}
\put(150,0){%
  \begin{picture}(85,90)(-5,0)
    \CArc(40,45)(15,0,360)
    \Photon(10,65)(35,50)24  \Vertex(35,50)1  \Text(8,68)[r]{$\phi_i$}
    \Line(25,45)(15,45)  \Vertex(25,45)1  \Text(13,45)[r]{${a_1}$}
    \Line(55,45)(65,45)  \Vertex(55,45)1  \Text(67,45)[l]{${a_n}$}
    \Line(40,60)(40,70)  \Vertex(40,60)1  \Text(40,78)[m]{${a_0}$}
    \Text(22,35)[m]{$\cdot$}  \Text(28,29)[m]{$\cdot$}
    \Text(58,35)[m]{$\cdot$}  \Text(52,29)[m]{$\cdot$}
    \Line(40,30)(35,20)  \Line(40,30)(45,20)  \Vertex(40,30)1
     \Text(35,12)[r]{${a_k}$}  \Text(45,12)[l]{${a_{l}}$}
  \end{picture}
}
\end{picture}
\end{center}
\caption{Factorisation of the BRST operator.}
\label{brst}
\end{figure}
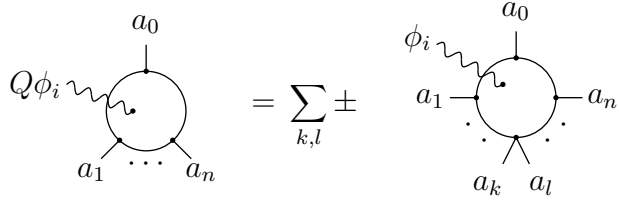

We want to view the mixed correlators as intertwiners between the closed 
membrane algebra and the deformations of the on-shell $L_\infty$ structure, 
given by the boundary correlators $G_{abc\ldots}$. An essential 
structure of the topological bulk theory is the BRST operator. 
A BRST operator acting on the closed string operator in the mixed 
correlators can be deformed to a contour around the boundary operators. 
Using the descent equations for the boundary operators gives the 
following identity, also depicted in \figref{brst}. 
\begin{eqnarray}\label{randtermen}
\Vev{ Q\phi_i\al_{a_0}\int\!\al_{a_1}^{(2)}\cdots
  \int\!\al_{a_n}^{(2)} }
&=& \Vev{ \phi_i\gerst{\al_{a_0},\al_{a_1}}\int\!\al_{a_2}^{(2)}\cdots
  \int\!\al_{a_n}^{(2)} } \nonumber\\
&&+ (-1)^{n+1}\Vev{ \phi_i\int\!\al_{a_1}^{(2)}\cdots
  \int\!\al_{a_{n-1}}^{(2)}\gerst{\al_{a_n},\al_{a_0}} } \\
&&+ \sum_{k=1}^{n-1} (-1)^{k}
 \Vev{ \phi_i\al_{a_0}\int\!\al_{a_1}^{(2)}\cdots
  \int\gerst{\al_{a_k},\al_{a_{k+1}}}^{(2)}\cdots\int\!\al_{a_n}^{(2)} }. 
\nonumber
\end{eqnarray}
In this derivation, the boundary operators are taken on-shell (BRST-closed), 
while for $\phi_i$ we take an arbitrary local membrane operator. 
The boundary terms in the factorised diagrams are related to points in 
the moduli space where two boundary operators approach each other. 
They arise from a total derivatives of the form  
$\int d\al_{a}^{(1)}=\int Q\al_{a}^{(2)}$. Its boundary term 
near another boundary operator will still contain 
a first descendant, which is integrated around the insertion point. 
Thus it involves the bracket rather than the product. We find that 
the bulk BRST operator corresponds to the operator $\delta_b$. 
More generally, if we include off-shell boundary operators we 
find that there are corrections from the boundary BRST operator, 
which are easily seen to correspond to the coboundary $\delta_Q$ 
acting on the maps. The coboundary operator on the deformation complex 
is therefore found to be $\delta_Q+\delta_b$, which is indeed the 
coboundary operator related to deformations of the DL (or more generally of 
the $L_\infty$) structure. 

\begin{figure}
\begin{center}
\begin{picture}(310,125)
\put(0,20){
  \begin{picture}(80,80)(0,10)
    \CArc(50,45)(15,0,360)
    \Photon(25,65)(43,50)24
    \LongArrowArc(45,45)(5,120,120)  \Text(25,70)[r]{$\phi_i^{(2)}$}
    \Photon(20,45)(45,45)24  \Vertex(45,45)1  \Text(17,45)[r]{$\phi_j$}
    \Line(33,27)(40,34)  \Vertex(40,34)1  \Text(28,22)[m]{$a_1$}
    \Line(67,27)(60,34)  \Vertex(60,34)1  \Text(65,22)[l]{$a_n$}
    \Line(50,60)(50,70)  \Vertex(50,60)1  \Text(50,77)[m]{$a_0$}
    \Text(50,25)[m]{$\cdots$}
  \end{picture}}
\put(95,55){$=\quad\displaystyle\sum_{k,l}\pm$}
\put(150,0){
  \begin{picture}(85,125)(-5,15)
    \CArc(40,100)(15,0,360)
    \Photon(15,125)(35,105)24  \Vertex(35,105)1  \Text(13,127)[r]{$\phi_i$}
    \Line(40,115)(40,125)  \Vertex(40,115)1  \Text(40,132)[m]{$a_0$}
    \Line(15,100)(25,100)  \Vertex(25,100)1  \Text(13,100)[r]{$a_1$}
    \Line(65,100)(55,100)  \Vertex(55,100)1  \Text(67,100)[l]{$a_n$}
    \Line(23,82)(30,89)    \Vertex(30,89)1   \Text(25,77)[r]{$a_k$}
    \Text(22,94)[r]{$\cdot$} \Text(24,89)[r]{$\cdot$}
    \Text(58,94)[l]{$\cdot$} \Text(56,89)[l]{$\cdot$}
    \Line(57,82)(50,89)  \Vertex(50,89)1  \Text(55,77)[l]{$a_{l}$}
    \Text(43,72)[l]{$b$}
    \CArc(40,45)(15,0,360)
    \Photon(10,45)(35,45)24  \Vertex(35,45)1  \Text(8,45)[r]{$\phi_j$}
    \Line(40,60)(40,85)  \Vertex(40,60)1  \Vertex(40,85)1
    \Line(23,27)(30,34)  \Vertex(30,34)1  \Text(30,22)[r]{$a_{k+1}$}
    \Line(57,27)(50,34)  \Vertex(50,34)1  \Text(55,22)[l]{$a_{l-1}$}
    \Text(40,25)[m]{$\cdots$}
  \end{picture}}
\put(255,50){$\pm\, (i\leftrightarrow j)$}
\end{picture}
\end{center}
\caption{Factorisation of the bracket.}
\label{brfact}
\end{figure}
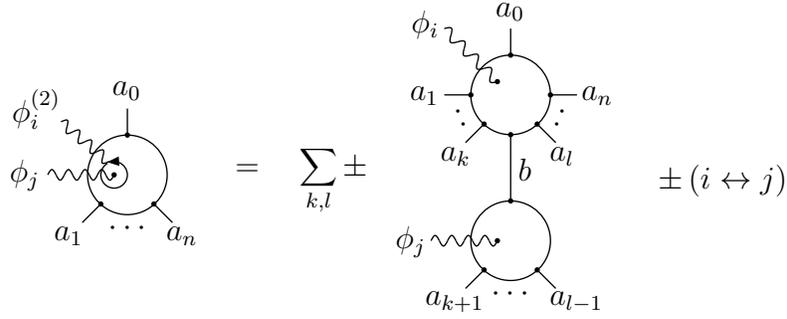

We can do the same with the inclusion of a second bulk operator, 
that is we look at the factorisation of the correlation function 
\begin{equation}\label{bigfact}
\Vev{\int\!(Q\phi_i)^{(3)}\int\!\phi_j^{(3)}\al_{a_0}\oint\!\al_{a_1}^{(1)}
\al_{a_2}\int\!\al_{a_3}^{(2)}\cdots\int\!\al_{a_n}^{(2)}}, 
\end{equation}
which vanishes on-shell. The basic difference is that the 
undeformed products $m$ (of any order) are replaced by the deformed products. 
Furthermore, there is an extra boundary term related to the two 
bulk operators coming close together. This involves the integral of 
the second descendant of $\phi_i$ around $\phi_j$, because of 
the total derivative term $d\phi_i^{(2)}$ coming from pulling
$Q$ through the descendants. It gives the bracket in the 
membrane theory. The vanishing of \eqref{bigfact}  gives the 
relation depicted in \figref{brfact},
\begin{equation}\label{haakje}
\Phi(\{\phi_i,\phi_j\}) = [\Phi(\phi_i),\Phi(\phi_j)]. 
\end{equation}

\begin{figure}
\begin{center}
\begin{picture}(400,100)
\put(0,10){
  \begin{picture}(95,80)(-15,10)
    \CArc(50,45)(15,0,360)
    \Photon(15,55)(45,45)25  \Vertex(45,45)1  \Vertex(15,55)1
     \Text(25,60)[l]{$\phi_k$}
    \Photon(15,55)(0,45)23  \Text(-3,45)[r]{$\phi_i$}
    \Photon(15,55)(5,70)23  \Text(3,72)[r]{$\phi_j$}
    \Line(33,27)(40,34)  \Vertex(40,34)1  \Text(35,22)[r]{${a_1}$}
    \Line(67,27)(60,34)  \Vertex(60,34)1  \Text(65,22)[l]{${a_n}$}
    \Line(50,60)(50,70)  \Vertex(50,60)1  \Text(50,77)[m]{${a_p}$}
    \Text(50,25)[m]{$\cdots$}
  \end{picture}}
\put(110,45){$=~\displaystyle\sum_{k}\pm$}
\put(165,0){
  \begin{picture}(170,110)(-5,10)
    \CArc(80,80)(5,0,360)
    \Line(80,85)(80,95)  \Vertex(80,85)1  \Text(80,102)[m]{${a_0}$}
    \Line(76,77)(50,56)  \Vertex(76,77)1  \Vertex(50,56)1
     \Text(58,70)[r]{$b$}
    \Line(84,77)(110,56)  \Vertex(84,77)1  \Vertex(110,56)1
     \Text(100,70)[l]{$c$}
    \CArc(40,45)(15,0,360)
    \Photon(10,55)(35,45)24  \Vertex(35,45)1  \Text(8,55)[r]{$\phi_i$}
    \Line(23,27)(30,34)  \Vertex(30,34)1  \Text(30,22)[r]{${a_{1}}$}
    \Line(57,27)(50,34)  \Vertex(50,34)1  \Text(55,22)[l]{${a_k}$}
    \Text(40,25)[m]{$\cdots$}
    \CArc(120,45)(15,0,360)
    \Photon(150,55)(125,45)24  \Vertex(125,45)1  \Text(153,55)[l]{$\phi_j$}
    \Line(103,27)(110,34)  \Vertex(110,34)1  \Text(110,22)[r]{${a_{k+1}}$}
    \Line(137,27)(130,34)  \Vertex(130,34)1  \Text(135,22)[l]{${a_{n}}$}
    \Text(120,25)[m]{$\cdots$}
  \end{picture}}
\put(350,45){$\pm\,(i\leftrightarrow j)$}
\end{picture}
\end{center}
\caption{Factorisation of the product.}
\label{prfact}
\end{figure}
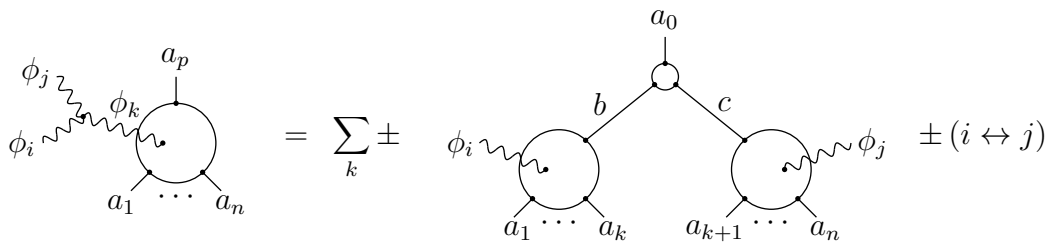

There is also a factorisation giving the bulk product as a 
boundary term, and several factorised correlation functions 
as the other boundary terms. However, it involves a 
codimension 3 boundary, starting from the deformed 
correlator with two deforming operators. This can be seen 
from the fact that we need to replace $\int\!\phi_i^{(3)}\int\!\phi_j^{(3)}$ 
by a single descendant $\int\!(\phi_i\cdot\phi_j)^{(3)}$. 
The factorisation is depicted in \figref{prfact}. 
From the fact that we have a codimension 3 boundary, 
it can be seen that the undeformed factor involves the bracket 
of the boundary theory. This gives the following identity:
\begin{equation}\label{produkt}
\Phi(\phi_i\cdot\phi_j)(\al_{a_1},\ldots,\al_{a_n}) 
= \sum_k\pm\{\Phi_i(\al_{a_1},\ldots,\al_{a_k}),\Phi_j(\al_{a_{k+1}},\ldots,\al_{a_n})\}.
\end{equation}

\section{The Topological Open Membrane}
\label{sec:tom}

In this section we discuss as an example an explicit topological open membrane 
theory. The model we will study is the membrane with only a WZ term, 
whose action is given by 
\begin{equation}\label{wz}
S = \int_M \frac{1}{6}C_{ijk}dX^i\wedge dX^j\wedge dX^k.
\end{equation}
This action appears for example as a suitable decoupling limit of the open 
supermembrane in M-theory \cite{bebe}. 
This action as it stands is quite singular for calculating correlation functions, 
as it is cubic. In order to allow ourselves to do calculations and quantise 
the action, we use a first-order formalism and BV quantisation techniques, 
as developed in \cite{js}. In this section we will only state the main points 
of the calculation and the final results, as it is just intended as a first 
example of the nontrivial deformation of the $DG$ structure. More worked-out 
calculations will appear in a forthcoming paper of one of the authors 
\cite{tom}.

\subsection{BV Quantisation of the Topological Open Membrane}

The explicit topological open membrane we will study is a 
BV-quantised membrane theory, which was discussed in \cite{js}. 
This theory is very much inspired by the Cataneo and Felder model (CF)
for the topological open string with a $B$-field WZ term \cite{cafe}. 
The easiest way to write down the CF model is to use superfields; these are 
functions on the worldsheet of bosonic coordinates $x^\mu$ and fermionic 
coordinates $\theta^\mu$. These superfields combine all the fields: 
physical fields, ghost fields and antifields. In the CF model there 
were two sets of superfields, which we will denote here\footnote{These 
superfields were called $\tilde X$ and $\tilde\eta$ in \cite{cafe}.} 
$\BX^i(x,\theta)$ and $\Bch_i(x,\theta)$ -- the first one bosonic, 
the second fermionic. They are the generating functionals of the scalars 
$X^i$ and $\chi_i$ and their descendants. This formalism can be viewed as a 
quantisation of the open string: the boundary operators are functions 
of the superfields $\BX^i$, while the superfields $\Bch_i$ play the role 
of ``momenta''. Moreover, together they generate the Hochschild 
cohomology of the open string algebra, e.g.\ $\Bch_i$ represents 
$\frac{\del}{\del\BX^i}\in\Hom(A,A)$. 

The explicit BV quantisation of the TOM theory defined by the WZ term 
goes very much along the same lines. We will not give an elaborate 
motivation, as this goes outside the scope of the present paper. 
Instead we will simply pose the model here, and give motivation for it later, 
by showing that the undeformed TOM is equivalent to the topological closed 
string theory given by CF. From the philosophy above, in order to construct 
the TOM we have to introduce two more sets of superfields, which we denote 
$\Bps^i$ and $\BF_i$, which serve as ``momenta'' for the two superfields 
$\BX^i$ and $\Bch_i$. The four superfields describing the TOM can be expanded as 
\begin{eqnarray*}
\BX^i &=& X^i + \rho^i_\mu\theta^\mu 
  +\frac12 X^i_{\mu\nu}\theta^\mu\theta^\nu  
  +\frac16 \rho^i_{\mu\nu\lambda}\theta^\mu\theta^\nu\theta^\lambda, 
 \\  
\Bch_i &=& \chi_i + H_{i\mu}\theta^\mu 
  +\frac12 \chi_{i\mu\nu}\theta^\mu\theta^\nu 
  +\frac16 H_{i\mu\nu\lambda}\theta^\mu\theta^\nu\theta^\lambda, 
 \\  
\Bps^i &=& \psi^i + A^i_\mu\theta^\mu 
  +\frac12 \psi^i_{\mu\nu}\theta^\mu\theta^\nu
  +\frac16 A^i_{\mu\nu\lambda}\theta^\mu\theta^\nu\theta^\lambda,
 \\  
\BF_i &=& F_i + \eta_{i\mu}\theta^\mu 
  +\frac12 F_{i\mu\nu}\theta^\mu\theta^\nu 
  +\frac16 \eta_{i\mu\nu\lambda}\theta^\mu\theta^\nu\theta^\lambda.
\end{eqnarray*}
These fields have ghost degree 0, 1, 1, and 2, respectively. The scalar 
components $(X^i,\chi_i,\psi^i,F_i)$ can be viewed as coordinates on the 
superspace $\Pi T(\Pi T^*M)$. Here $\Pi$ is an operator that shifts the 
degree in the fibre by one. Viewing $(x^\mu,\theta^\mu)$ as coordinates 
on the superspace $\Pi TN$, where 
$N$ is the worldvolume of the membrane, these fields can be viewed as 
parametrising a map $\Pi TN\to \Pi T(\Pi T^*M)$ between the two superspaces. 
We will choose boundary conditions such that the new fields $\Bps^i$ and $\BF_i$ 
vanish on $\del N$. This means that the boundary $\del N$ maps to the base space 
$\Pi T^*M$ of the target space. 

In order to get a BV quantisation, we need to introduce a BV (anti)bracket. 
From our motivation of choosing $\BF$ and $\Bps$ as the ``momenta'' of 
the superfields $\BX$ and $\Bch$, we have a natural symplectic structure 
on the superfields above, 
\begin{equation}
\Bom_{BV} = \int_N\dint{d^3\theta}\Bigl(\delta\BX^i\delta\BF_i 
  + \delta\Bch_i\delta\Bps^i\Bigr), 
\end{equation}
where $\delta$ denotes the $d$-operator (De Rham differential) on field space. 
This is a symplectic form of ghost degree $-1$. This symplectic structure 
defines the BV bracket, which is dual to it, and can formally be written 
\begin{equation}
(\cdot,\cdot)_{BV} = \dd{\BX^i}\wedge\dd{\BF_i} + \dd{\Bch_i}\wedge\dd{\Bps^i}.
\end{equation}
This is easily seen to derive from a BV operator $\triangle$. 

Motivated by CF, we will write down an undeformed membrane theory using a 
Poisson bivector $b^{ij}$ on $M$, i.e.\ $b^{il}\del_l b^{jk}+perms.=0$. 
The BV action we propose is given by 
\begin{equation}
S_0 = \int_N\dint{d^3\theta} \Bigl( \BF_i D\BX^i + \Bps^i D\Bch_i + \BF_i\Bps^i 
 + \Bb^{ij}\BF_i\Bch_j +\frac{1}{2}\del_k\Bb^{ij}\Bps^k\Bch_i\Bch_j \Bigr), 
\end{equation}
where $D=\theta^\mu\del_\mu$, and $\Bb^{ij}$ denotes the pull-back by 
the superfield $\BX$ to $\Pi TN$, $\Bb^{ij}(x,\theta)=b^{ij}(\BX(x,\theta))$. 
It is easily seen that the BV action above satisfies both the classical and 
the quantum master equation, $\triangle S_0=(S_0,S_0)_{BV}=0$. 
The BRST operator is determined through 
$\BQ_0=(S_0,\cdot)_{BV}$. Because the auxiliary fields $\BF_i$ appear only linearly, 
we can exactly integrate them out. After solving for $\Bps_i$ in this equation, 
the action reduces to a pure boundary term
\begin{equation}
S_{CF} = \int_{\del N}\dint{d^2\theta} \Bigl( 
 \Bch_iD\BX^i +\frac{1}{2}\Bb^{ij}\Bch_i\Bch_j\Bigr)
 = \int_{\del N} \Bigl( H_i dX^i + \chi_i d\rho^i 
 + \frac{1}{2}b^{ij}H_iH_j +\cdots\Bigr). 
\end{equation}
This is precisely the action of the Cattaneo-Felder model, as announced. 
This is related to the usual topological closed string with just the 
$B$-field WZ term by integrating out $H$. 

The boundary operators are determined by functions $f$ of the scalar fields 
$X^i$ and $\chi_i$, that is functions on the base space $\Pi T^*M$ of the 
target space. The corresponding boundary operator $\al_f$ and its descendants 
combined as 
\begin{equation}
\al_f+\theta\al_f^{(1)}+\frac12\theta^2\al_f^{(2)}=f(\BX,\Bch).
\end{equation}
We will sometimes denote this by $\Bf$. It is natural to view the 
space of functions on $\Pi T^*M$ as the polynomial algebra 
$\C[\{X^i\},\{\chi_i\}]$ generated by $X^i$ and $\chi_i$. 
By formally replacing the fermionic generators $\chi_i$ by 
the basic vector fields $\del_i$, one sees that the boundary 
operators are in one-to-one correspondence with the 
multi-vector fields on the target space $M$. Our undeformed 
boundary algebra $A$ will thus be the algebra of 
multi-vector fields $A=\Gamma(M,\ext[*]TM)$. 

The three-point functions determine a structure of an algebra on these 
boundary operators, which indeed turns out to be a 2-algebra. 
More precisely, for the product and the bracket this relation will be given by 
\begin{equation}
\Vev{\al_\delta\al_f\al_g} \equiv \Vev{\al_\delta\al_{f\cdot g}},\qquad 
\Vev{\al_\delta\oint\!\al_f^{(1)}\al_g} \equiv \Vev{\al_\delta\al_{\{f,g\}}},
\end{equation}
where we took the outgoing state corresponding to a $\delta$-function on the 
target space. 
The product is easily seen to be the wedge product on the multi-vector fields. 
The bracket is given by 
\begin{equation}
\{f,g\} = \frac{\del f}{\del X^i}\frac{\del  g}{\del\chi_i} 
 + (-1)^{|f|}\frac{\del f}{\del\chi_i}\frac{\del g}{\del X^i} .
\end{equation}
This is a consequence of the $\Bps^i\BF_i$ term in the BV action. 
As a bracket on the multi-vector fields, this bracket is well-known in mathematics. 
It is called the Schouten-Nijenhuis bracket. It is the unique extension 
to the full algebra of multi-vector fields of the Lie bracket 
on vector fields. 

When $b$ is nonzero, there is also a differential, so that the boundary string 
theory is a differential Gerstenhaber algebra. 
This differential is given by the BRST operator restricted to 
the boundary,
\begin{equation}
Qf = b^{ij}\chi_j\frac{\del f}{\del X^i} 
 + \frac{1}{2}\del_ib^{jk}\chi_j\chi_k\frac{\del f}{\del\chi_i}.
\end{equation}
It is easily checked that $Q$ is nilpotent if $b^{ij}$ is a 
Poisson structure. 

We conclude that the undeformed topological open membrane 
we proposed above is given by the algebra of multi-vector fields. 
It is supplied with the differential $Q$ above, the wedge product and the 
Schouten-Nijenhuis bracket, which indeed makes it into a 2-algebra. 
Our next task is to study the deformation of this 2-algebra. 
We first propose a natural deformation in the context of our 
BV-quantised theory.

\subsection{Deformations of the TOM}

The boundary string theory will be deformed by coupling the TOM to a 
bulk operator. We can construct bulk operators corresponding to 
functions $f(X,\chi,\psi,F)$ on the full target space $\Pi T(\Pi T^*M)$. 
They are given by the pull back to the worldvolume $\Bf=f(\BX,\Bch,\Bps,\BF)$, 
using the superfields. This generates all descendants of 
the operator and to conserve ghost number, this should have degree 3. 
The natural topological deformation is to turn on a 
3-form deformation in the open membrane theory. 
Given a 3-form $c$ this defines an operator $\phi_c$, which for $b=0$ is 
given by 
\begin{equation}
\int_N \phi_c^{(3)} = \int_N\dint{d^3\theta} \frac16\Bc_{ijk}\Bps^i\Bps^j\Bps^k.
\end{equation}
We will use this operator as the deformation of the BV action functional. 
It will turn out that the $b$-field in general does not have 
to define a strict Poisson structure in the deformed case, 
so we will for now drop this requirement. 
The totally deformed BV action, including $b$, becomes \cite{js} 
\begin{eqnarray}\label{bvaction}
S &=& \int_N\dint{d^3\theta} \Bigl(
\BF_i D\BX^i + \Bps^i D\Bch_i + \BF_i\Bps^i 
 + \Bb^{ij}\BF_i\Bch_j +\frac{1}{2}\del_k\Bb^{ij}\Bps^k\Bch_i\Bch_j 
+\frac{1}{2}\Bb^{il}\del_l\Bb^{jk}\Bch_i\Bch_j\Bch_k \nonumber\\
&&\phantom{\int_N\dint{d^3\theta} \Bigl(}
 + \frac{1}{6}\Bc_{ijk}(\Bps^i+\Bb^{il}\Bch_l)(\Bps^j+\Bb^{jm}\Bch_m)(\Bps^k+\Bb^{kn}\Bch_n)
\Bigr). 
\end{eqnarray}
This action functional satisfies the BV master equation if the total 
field strength given by 
\begin{equation}
\Bh^{ijk} = \Bb^{il}\del_l\Bb^{jk} + \Bb^{jl}\del_l\Bb^{ki} + \Bb^{kl}\del_l\Bb^{ij}
 + \Bb^{il}\Bb^{jm}\Bb^{kn}\Bc_{lmn},
\end{equation}
vanishes. Notice that this implies that $b^{ij}$ is not necessarily a Poisson structure. 

If we now integrate out $\BF$, the second line in \eqref{bvaction} reduces to the WZ 
term \eqref{wz} of the $c$-field. This motivates our choice for the deforming operator, 
and for the whole model, since it shows that the  model serves as a well-defined 
quantum action for the ill-defined theory based on the WZ term. 

To calculate the first-order corrections to the algebraic structure we need 
to calculate the corresponding correlation functions, which define the map 
$\Phi_c$ corresponding to the operator $\phi_c$. This can be related to 
a deformation on the algebra of multi-vector fields. For example, 
we can write
\begin{equation}
\Phi_c(\al_f,\al_g) \equiv \al_{\{f,g\}_1}, 
\end{equation}
where $\{\cdot,\cdot\}_1$ is the first-order deformation of the 
bracket on the multi-vector fields. 
In the next subsection we will use the Hochschild complex to 
calculate the effect on the algebra, at least in a first-order quantisation. 
The field theory we now have can in principle be used to  calculate 
the correspondence of the Hochschild cohomology -- the deforming operators -- 
and the Hochschild complex -- the differential operators -- as a perturbation 
series in $c$ (formality).

\subsection{Hochschild Cohomology of the 2-Algebra of Multi-Vector Fields}

In \secref{sec:defcplx}, we saw that the possible deformers are essentially 
given by elements of the Hochschild cohomology. 
We will now calculate this cohomology for the topological open membrane 
theory. We start with the situation $b=0$. 

We saw that the operators of the boundary closed string form
the algebra of functions on $\Pi T^*M$, which we represent by the algebra of 
polynomials $A=\C[\{X^i\},\{\chi_i\}]$. As explained above, this 
corresponds to the algebra of multi-vector fields $\Gamma(M,\ext[*]TM)$. 
This is naturally a graded algebra, with 
the degree corresponding to the vector degree. This means that the generators 
$X^i$ have degree 0, and $\chi_i$ have degree 1. This algebra indeed has the 
structure of a Gerstenhaber algebra or 2-algebra, with the product $m$ given 
by the wedge product and the bracket $b$ given by the Schouten-Nijenhuis 
bracket, defined by 
\begin{equation}
\{\al,\beta\} = \frac{\del\al}{\del X^i}\frac{\del\beta}{\del\chi_i}
 +(-1)^{|\al|}\frac{\del\al}{\del\chi_i}\frac{\del\beta}{\del  X^i} .
\end{equation}
This is about the simplest nontrivial 2-algebra one can construct. 

The deformation of the Gerstenhaber algebra of multi-vector fields is determined 
by the Hochschild cohomology. The Hochschild complex is given by the 
algebra of multilinear operators acting on the algebra $A$, 
$\Hoch(A)=\bigoplus_{n}\Hom(A^{\otimes n},A)$. This is the algebra of 
multi-differential operators. On the cohomology act the 3 differential 
described above. 
Taking the partial cohomology with respect to the differential $\delta_m$ associated 
to the ordinary product, we can describe these multi-differential operators by 
introducing anticommuting coordinates $\psi^i$, representing $\del_{\chi_i}$, and 
commuting variables $F_i$, representing $\del_{X^i}$. The Hochschild cohomology 
can be described as a polynomial algebra: 
$H^*_{\delta_m}\bigl(\Hoch(A)\bigr)=\C[\{X^i\},\{\chi_i\},\{\psi^i\},\{F_i\}]$, 
see \appref{app:hochpol}. The degree of the generators $\psi^i$ is 1, while the 
degree of $F_i$ should be taken 2.\footnote{These correspond precisely to the extra 
fields in the BV action. This correspondence can in fact be taken quite seriously.}
There is still a differential left, related to the bracket. It is defined in a similar 
way to the Gerstenhaber differential, but with the product replaced by the bracket. 
This differential is easily calculated on the above polynomial algebra to be given by 
\begin{equation}
\delta_b = \psi^i\frac{\del}{\del X^i}+F_i\frac{\del}{\del\chi_i}, 
\end{equation}
which correctly has degree 1. The full Hochschild cohomology is now the cohomology 
of the above polynomial algebra with respect to this differential. This algebra 
has a natural Poisson structure of degree $-2$, given by 
\begin{equation}
\{\al,\beta\} = \frac{\del\al}{\del X^i}\frac{\del\beta}{\del F_i} 
 -(-1)^{|\al|} \frac{\del\al}{\del\chi_i}\frac{\del\beta}{\del\psi^i} 
\pm(\al\leftrightarrow\beta).
\end{equation}
The structure of the differential $\delta_b$, the bracket of degree $-2$ 
and the product makes the Hochschild cohomology into a 3-algebra, 
which is just a differential Poisson algebra, except from the degree of the bracket. 

The cohomology with respect to the differential $\delta_b$ removes all dependence on 
$\chi$ and $F$, so that in the end we are left with only polynomials of $X^i$ and $\psi^i$. 
Hence the cohomology equals that of the differential forms on $\R^n$. 
Note that the Poisson bracket of the 3-algebra is identically zero on the 
cohomology. 

In general, for $A=\Gamma(M,\ext[*]TM)$, we find  $H^*\Hoch(A)=H^*(M)$. 
This means that for sufficiently large $p$, we have $H^p(\Def(A))=H^{p+2}(M)$. 
Especially, $H^1(\Def(A))=H^3(M)$. This term in the complex determines the 
actual deformations. The element in the Hochschild cohomology
corresponding to a closed 3-form $c$ is represented by the 
polynomial $\frac{1}{6}c_{ijk}(X)\psi^i\psi^j\psi^k$. We are of course 
interested in the corresponding element in the full Hochschild complex, 
that is the map $\Phi_c$ deforming the algebra. To find it, remember that 
$\psi^i$ corresponds to the operator $\frac{\del}{\del\chi_i}$ in the complex. 
This corresponds to a naive canonical quantisation, which gives 
\begin{equation}
\frac{1}{6}c_{ijk}(X)\frac{\del}{\del\chi_i}\wedge\frac{\del}{\del\chi_j}\wedge\frac{\del}{\del\chi_k}. 
\end{equation}
Of course, this is only the leading term in the map from the Hochschild 
cohomology to the complex.\footnote{It should be compared to the 
leading term $\theta^{ij}\del_i\wedge\del_j$ for the deformation of the product 
in noncommutative geometry.} 
Notice that this is a trilinear differential operator. This means that a 
trilinear product in the $L_\infty$ algebra is deformed. 

Let us now turn on a $b$-field $b^{ij}$, which we will take constant for 
simplicity. This introduces a derivation $Q$ on the algebra, and the 
calculation of the cohomology for the double complex is more complicated, 
as we now have two coboundary operators $\delta_Q$ and $\delta_b$ on the complex. 
The total coboundary operator on the double complex $C=\Hoch(A)$ is given by 
$D=d+\delta=\delta_b+\delta_Q$. With both differentials 
nonzero, we can in general calculate the cohomology using spectral sequence 
techniques, see \appref{app:dcplx}. This basically amounts to solving a 
series of descent equations. Starting from a class $d$-closed element 
$\al_0$, we have descent equations $\delta\al_0=-d\al_1$, etcetera. 
The two coboundary operators on the double complex are given by 
\begin{equation}
\delta \equiv \delta_Q = b^{ij}\chi_j\dd{X^i} - b^{ij}F_j\dd{\psi^i},\qquad
d \equiv \delta_b = \psi^i\dd{X^i} + F_i\dd{\chi_i}. 
\end{equation}
It turns out that the descent equations can be solved introducing the following 
operator
\begin{equation}
\gamma = b^{ij}\chi_j\frac{\del}{\del\psi^i}.
\end{equation}
It is easily checked that $[d,\gamma]=-\delta$. This can be used 
to solve $\al_1=\gamma\al_0$, $\al_2=\gamma\al_1$, and so on. 

Let us see what this implies for the deformation term, when we turn $b$ on. 
First note that the operator $d$ is not affected by turning on $b$, 
therefore we still conclude that the $d$-cohomology class $\al_0$ is represented by an 
element $\frac{1}{p!}\al_{i_1\ldots i_p}(X)\psi^{i_1}\cdots\psi^{i_p}$, 
where $\al_{i_1\ldots i_p}(X)$ is a closed $p$-form. The effect of $\gamma$ 
is to replace $\psi^i$ by $b^{ij}\chi_j$. Therefore, the total 
class $\al$ is given in terms of the same form, but with $\psi^i$ 
replaced by $\psi^i+b^{ij}\chi_j$, 
\begin{equation}
\al = \frac{1}{p!}\al_{i_1\ldots i_p}(X)(\psi^{i_1}+b^{i_1j_1}\chi_{j_1})
 \cdots(\psi^{i_p}+b^{i_pj_p}\chi_{j_p}). 
\end{equation}
Most interestingly, the class in the third cohomology related to the closed 3-form 
$c$ is given by 
\begin{equation}
\frac{1}{6}c_{ijk}\psi^i\psi^j\psi^k
+ \frac{1}{2}c_{ijk}b^{il}\chi_l\psi^j\psi^k
+ \frac{1}{2}c_{ijk}b^{il}b^{jm}\chi_l\chi_m\psi^k
+ \frac{1}{6}c_{ijk}b^{il}b^{jm}b^{kn}\chi_l\chi_m\chi_n.
\end{equation}
This corresponds precisely to the deformation term in the action 
\eqref{bvaction}. 

Using the first-order map (the ``quantisation'') from the cohomology to 
the Hochschild complex, this translates into the following set 
of deformed operations in the algebra:
\begin{eqnarray}\label{defTOM}
Q &=& b^{ij}\chi_j\frac{\del}{\del X^i} 
 + \frac{1}{2}(\del_kb^{ij}+c_{klm}b^{li}b^{mj})\chi_i\chi_j\frac{\del}{\del\chi_k}
 + \CO(c^2), 
  \nonumber \\
\{\cdot,\cdot\} &=& \frac{\del}{\del X^i}\wedge\frac{\del}{\del\chi_i} 
 + \frac{1}{2}c_{ijk}b^{kl}\chi_l\frac{\del}{\del\chi_i}\wedge\frac{\del}{\del\chi_j} +\CO(c^2), 
  \\
\{\cdot,\cdot,\cdot\} \equiv b_3 &=&
  \frac{1}{6}c_{ijk}\frac{\del}{\del\chi_i}\wedge\frac{\del}{\del\chi_j}\wedge\frac{\del}{\del\chi_k}
 + \CO(c^2).
  \nonumber
\end{eqnarray}
The corrections to the BRST operator $Q$ and the bracket $\{\cdot,\cdot\}$ 
are precisely given by the corrections in the higher terms of the spectral sequence: 
the sum is simply the quantisation of the total representative. 
These operations satisfy the relations of a ``$L_3$ algebra''. 
Together with the undeformed product, it satisfies the relations of a ``$G_3$ algebra''. 

More precisely, the above operations should be calculated by computations of the 
corresponding membrane correlators. Indeed, direct tree-level computations 
confirm the naive quantisation rules \cite{tom} to this order in $c$. 
More generally, higher order corrections to these operations can be given by 
loop calculations in the TOM.

\subsection{Effective Target Space Action}

We will  comment briefly on the consequence of the deformations we found. 

The correlation functions determine an effective action in the target 
space $M$, which is defined as the generating functional  
of the correlation functions of the boundary operators. As we saw, the 
boundary operators are related to functions of $X$ and $\chi$, which 
can be identified with multi-vector fields. The physical fields in the 
effective action correspond to the physical boundary operators in the open 
membrane theory. These are the operators of ghost degree 2 
$\CB=\frac{1}{2}B^{ij}(X)\chi_i\chi_j$, which correspond to degree 2 
multi-vector fields. Interpreting the effective action as the generating 
functional of the correlation functions $F_{a_0\ldots a_n}$ of the open 
membrane theory gives in general an effective action functional which 
to first order in $c$ can be written in the form 
\begin{equation}
S_{eff} = \int_{\Pi T^*M} \biggl(\frac{1}{2}\CB\cdot Q\CB 
 + \frac{1}{3}\CB\cdot\gerst{\CB,\CB} + \frac{1}{4}\CB\cdot\gerst{\CB,\CB,\CB}\biggr),
\end{equation}
where we integrate over the zero-modes of $X^i$ and $\chi_i$. Precisely 
such a form for the action of the closed string field theory was proposed 
by Zwiebach \cite{zwie} for the bosonic closed string, which was shown to 
satisfy the (quantum) master equation. Generalising his proposal for more 
general closed string field theories, this is of course what it reduces to 
in the case of the TOM. The integration over $\chi$ picks out the top 
component in terms of the multi-vector degree, which is nonzero only for $D=5$. 
In other dimensions, we cannot consistently truncate to the physical degrees of 
freedom, and we also have to take into account other non-physical modes. 
It seems that 5 dimensions is very natural for this action. In this situation 
the action is an interacting topological field theory which is very reminiscent 
of Chern-Simons, but with a 2-form gauge field. This is closely related to the 
way the Chern-Simons action arises in topological open string theory \cite{witcs}, 
which is exactly the analogue for the open string derivation we gave here 
for the open membrane. Notice that this theory is already interacting 
for $c=0$, as we still have a cubic term coming from the bracket. 
The $c$-field gives a further quartic interaction term. 

We can indeed interpret much of the deformation theory in terms of a 
generalised gauge theory. Let us first go to a representation in terms 
of differential forms rather than multi-vector fields. This can be done 
if we take as a background an invertible $b^{ij}$, and write the algebra 
$A$ in terms of $\chi^i=b^{ij}\chi_j$. Indeed functions of $X^i$ and $\chi^i$ 
can be identified with differential forms, if we identify $\chi^i=dX^i$. 
In this identification, the BRST operator $Q$, for $c=0$, is identified 
with the De Rham differential. 

Turning on a boundary operator $B=\frac{1}{2}B^{ij}\chi_i\chi_j$ affects $Q$. 
The perturbed BRST operator has the form
\begin{equation}
Q_B = Q+\{B,\cdot\}, 
\end{equation}
in terms of the bracket on the algebra of multi-vector fields. This can be 
interpreted as a covariant $d$-operator. Let us now consider what happens if 
we start from a nonzero $c$. The unperturbed BRST operator $Q$ has a 
connection part proportional to $c$, as well as deformed bilinear and 
trilinear brackets, as can be seen form \eqref{defTOM}. 
Now if we turn on a 2-form $B$, the abstract formula for the deformed 
BRST operator $Q_B$ is slightly changed due to the presence of the trilinear 
product,
\begin{equation}
Q_B = Q+\{B,\cdot\} +\frac12 \{B,B,\cdot\}. 
\end{equation}
Moreover, we also find a correction for the bracket proportional to 
the trilinear bracket, 
\begin{equation}
\{\cdot,\cdot\}_B = \{\cdot,\cdot\} + \{B,\cdot,\cdot\}.
\end{equation}
We might interpret this as a covariant bracket. 
We can repeat much of what we know about gauge theory to 
this 2-form theory. There is a field strength given by 
\begin{equation}
H =  QB+\{B,B\}+\{B,B,B\}. 
\end{equation}
The equations of motion for the above Chern-Simons like theory 
require this field strength to vanish. Also, we have gauge invariances of the form
\begin{equation}
\delta_\Lambda B = Q_B\Lambda = Q\Lambda+\{B,\Lambda\}+\{B,B,\Lambda\}.
\end{equation}
The field strength $H$ is  gauge covariant in the sense that 
$\delta_\Lambda H = \{H,\Lambda\}_B$.
Note that the gauge transformation of $H$ involves the covariant bracket.

\section{Conclusions and Outlook}
\label{sec:concl}

We studied deformations of (topological) closed string theories from
a worldsheet point of view. We saw that on-shell closed string
theories have the structure of a Gerstenhaber algebra (2-algebra),
which  generalises off-shell to a homotopy Gerstenhaber or $G_\infty$
algebra. Deformations of the string theory can therefore 
mathematically be described by a deformation of these Gerstenhaber
structures. Deformations of algebras are in general encoded by
the deformation complex, whose essential ingredient is the Hochschild
complex. We demonstrated how the structure of the Hochschild complex can
be read off from the deformations of the correlation functions of the
string theory. In particular, this shows the algebraic structure of
the Hochschild complex to be either a Poisson algebra or a 
Gerstenhaber algebra. 

We found that in principle one can write down three different
deformation complexes for the same vector space of operations. 
They correspond to the deformation of different operations in the 
closed string theory. In particular, the deformation of the closed 
string by itself deforms only the (homotopy) associative part of the 
(homotopy) Gerstenhaber algebra. On the other hand, closed strings 
that arise as the boundary theory of a topological open membrane show 
the deformation structure of the (homotopy) Lie algebra. A relation to
AdS$_3$/CFT$_2$ may be established here, since the membrane as a
deformation of the CFT on the boundary is exactly the AdS/CFT
correspondence.

Whether the third possible deformation, that of the Gerstenhaber
structure, has a physical interpretation remains an open question. On
the one hand it appears to be related to a string theory, on the other
hand ghost number conservation is probably violated. This may indicate 
of a breakdown of conformal invariance, which tempts to speculate about 
a relation with the $(1,1)$ LST.

We saw that we could define deformations for only two of the three 
basic operations in the closed string theory at the same time. 
It is not completely clear what can happen to the third operation 
when we deform the other two. In some cases (WDVV) it is undeformed. 
In other cases however the structure may get lost.

For open strings, the $A_\infty$ structure determines a superpotential
on the moduli space. The higher structure constants therefore give
obstructions to the flat directions due to the higher order
contributions to the superpotential. This raises the question whether
in the closed string case there are similar situations, where the
closed string higher structure constants give nontrivial
superpotentials and therefore higher obstructions. As of yet, 
there are no known examples of such phenomena in the physics literature, 
making it unclear if we really need the full $A_\infty$ structure in general. 

Mathematically speaking, the topological open membrane describes the
deformation of the algebra of multi-vector fields. A nontrivial third
homotopy is found in the Lie substructure. How will this be in the
generality of the operad formulation of Kontsevich \cite{kon2}, which
can be seen as a mold for describing deformations of extended objects
in string theory? In this context it is also particularly interesting
to examine more closely the relation between our treatment and the
geometric one of \cite{laza}, in which two-dimensional open-closed
field theories with very general boundary data are approached
axiomatically.

The precise relation to little string theory, $(2,0)$ CFT, and
M5-branes remains to be studied. The effective theory we wrote down 
seems more natural in 5 dimensions rather than in 6, which might indicate 
some relation to D4-branes. We may wonder if the relation $b$ to 
2-form gauge field and $c$ to a 3-form background is valid on the nose. 
The deformation for M5-branes should be related to the total field
strength $H=dB+C$. For the TOM the ``field strength'' $h$ is
constrained to vanish, while the field $c$ seems to deform the algebra
(or rather $b$ and $c$ combined). A related question is the choice of 
boundary conditions for the fields. In the last section we chose 
$\psi$ to vanish at the boundary, leading naturally to multi-vector 
fields for the boundary operators. One can in fact also choose 
Dirichlet boundary conditions for $\chi$ instead of $\psi$. This leads 
to boundary operators naturally induced by differential forms. Whether any 
of these choices, or perhaps both, corresponds to a decoupling limit 
of M5-branes is a question for further research. 

Another interesting connection can be found by relating to mathematics. 
The deformed $L_\infty$ algebra of the TOM that we found, including the 
trilinear bracket, can be seen to be the structure of a Courant algebroid 
\cite{cour,royt,wein}. This is a certain fibred generalisation of a 
quasi-Hopf algebra (quantum group), which arose in the study of
constrained quantisation. More precisely, the structure we found 
in the TOM is that of an exact Courant algebroid. In general, 
exact Courant algebroids are characterised by an element of $H^3(M,\R)$. 
In our language, this corresponds to the deformation $c$. The 
construction of this class is rather analogous to the class in 
$H^2(M,\R)$ of a ``local line bundle'' (more precisely, an algebroid 
of the form $TM\oplus\R$). When this second cohomology class is an 
integral class, this can be extended to a genuine global line bundle. 
The meaning of integrality for the third cohomology class is still 
mysterious, and is related to a global object for the Courant algebroid. 
Suggestions have been made that this should be a gerbe. The relation 
of the TOM to 2-form gauge theories indeed is very suggestive in that direction. 
One of the authors is currently involved in further investigations 
along these lines \cite{hofpar}.

The algebraic structure of the deformed TOM could also be helpful in finding 
a ``nonabelian'' generalisation of 2-form gauge theories. String theory suggests 
the existence of these theories in connection with multiple M5-branes. 
In the case of D-branes, the structure of the noncommutative gauge theory 
related to deformed open strings and the nonabelian gauge theory related to 
multiple D-branes is very similar. Analogously, we could expect the structure of 
multiple M5-branes and deformed M5-branes to be similar in an appropriate sense. 
There exist more general Courant algebroids which combine the nonabelian 
structure of Hopf algebras and the fibration structure of the deformed 
tangent space we found in the TOM. This is also very suggestive for 
a generalisation. 

The 2-form CS theory we found as an effective theory of the TOM in the
target space can be used to describe moduli spaces of flat 2-form
theories. If the speculation above turns out to be correct, this can
be interpreted as the moduli space of flat gerbes.

\ack

We like to thank Erik Verlinde and Robbert Dijkgraaf for useful
discussions and remarks. We especially like to thank Jae-Suk Park 
for a preliminary view of his work, and for discussions. 
We are grateful to the Spinoza Institute at Utrecht University and 
the Dept. of Physics at Princeton University where part of this research was performed. 
This research was partially supported by DOE grant \# DE-FG02-96ER40559. 

\appendix

\section{Algebras up to Homotopy}
\label{app:hom}

A \emph{homotopy associative} or \emph{$A_\infty$-algebra} can be defined in terms 
of a derivation $d$ acting on the tensor algebra $\CT A=\bigoplus_{n\geq0}A^{\otimes n}$ 
of a (graded) vector space $A$. The derivation is completely determined by 
the map from $\CT A$ to $A$. We denote the 
component of $d$ mapping the $n$th tensor product $A^{\otimes n}$ 
to $A$ by $d_n$. So we have $d=d_1+d_2+d_3+\cdots$. 
All $d_k$ are derivations in the sense that
\begin{equation}
d_k(a_1, \ldots, a_{k+n}) = \sum_{i=0}^{n} 
 (-1)^{i(k-1)}(a_1, \ldots, d_k(a_{i+1}, \ldots, a_{i+k}), \ldots, a_{k+n}).
\end{equation}
Furthermore, $d$ is a twisted differential in the following sense. 
Considering the shifted algebra $\Pi A=A[1]$.\footnote{For an integer $k$, 
$[k]$ denotes a shift of the degree of a complex $C =\bigoplus_n C^n$ by $k$, 
that is $C[k]^n := C^{k+n}$; therefore, $\Pi C^{n}=C^{n+1}$. 
Physically, the shift corresponds to descent.}
The shifted maps $\tilde d_k=\Pi\circ d_k\circ (\Pi\inv)^{\otimes k}$ should form 
a coboundary on the shifted algebra, i.e.\ $\tilde d^2=0$, of degree $1$. 
This implies an infinite number of homogeneous relations for the $d_k$: 
for any $n\geq0$, 
\begin{equation}
\sum_{k+l=n+1}(-1)^{(k-1)l}d_k\circ d_l=0.
\end{equation}
The map $d_k$ has degree $2-k$. 
Explicitly, the first few relations read $d_1^2 = 0$, $d_1d_2 =
d_2d_1$, $d_2^2 = -d_1d_3-d_3d_1$, $d_2d_3-d_3d_2 =
-d_1d_4-d_4d_1$. These say that $d_1$ is a differential on $A$, $d_2$
is a product for which $d_1$ is a derivation, $d_3$ gives a correction
to the associativity of this product ($d_2^2$ is the associator), etc. 

\emph{Homotopy Lie} or \emph{$L_\infty$-algebras} are defined in a similar way.  
We also start with a (graded) space $A$. The only difference is
that everything should be (graded) anti-symmetric; the
tensor product of the algebra is replaced by the (graded) exterior
product, $\bigoplus_n\ext[n]A$, and the products $d_n$ are all
(graded) anti-commutative. More precisely, the differential $d=\sum_n d_n$ 
can be considered as an operator $\tilde d$ operating on 
$\bigoplus_n S^n(A[1])\simeq\bigoplus_n(\ext[n]A)[n]$ \footnote{This relation is induced by 
$(\Pi a_1\otimes\cdots\otimes\Pi a_n)_S\to(-1)^{\sum_k(n-k)|a_k|} a_1\wedge\cdots\wedge a_n$.} 
by conjugating with the 
shift, such that $\tilde d^2=0$. They are called brackets; for example $d_2^2=0$ is
the Jacobi identity for the Lie bracket defined by $d_2$.

A \emph{Gerstenhaber algebra} ($G$-algebra) is a $\mathbb{Z}$-graded
algebra with a graded commutative associative product $\cdot$ of degree 0 and
a bracket $[\cdot,\cdot]$ of degree $-1$ (the Gerstenhaber bracket), which is such 
that $A[1]$ is a graded Lie algebra. Furthermore, the map $[\alpha,\cdot]$ must 
be a graded derivation of the product, 
\begin{equation}
  [\alpha,\beta\cdot\gamma] = [\alpha,\beta]\cdot\gamma
+ (-1)^{(|\alpha|-1)|\beta|}\beta\cdot [\alpha,\gamma]. 
\end{equation}
We can generalise this to a \emph{differential Gerstenhaber algebra} (or $DG$)
by adding a differential $\delta$ of degree 1, satisfying the graded derivation 
conditions with respect to the product and the bracket. Note that the shift 
$A[1]$ of a $(D)G$ algebra has the structure of a $L_\infty$ algebra. 
Hence there is a degree one differential on $S^*(A[2])$ which squares to zero. 

There does not seem to be an overall agreement over the notion of 
\emph{homotopy Gerstenhaber algebra} or \emph{$G_\infty$-algebra} in 
the literature. Some possible definitions are given in \cite{tamtsy,kvz,getzjon2}. 
They are fairly complicated constructions, and we will not attempt to 
give a definition here. We will mainly observe that they contain at least 
an $A_\infty$ and a $L_\infty$ subalgebra, with a shared differential.

\section{The Hochschild Cohomology of a Polynomial Algebra}
\label{app:hochpol}

We can give an explicit description of the Hochschild cohomology of a
general polynomial algebra. Consider the algebra of polynomials in a
finite number of $\Z$-graded variables $x^i$ of degree
$\deg(x^i)=q_i\in\Z$, so the space $A=\C[x^1,\ldots,x^N]$. We view it
as an algebra over the operad $H_*(C_d)$ (see \cite{kon2}) so a
$d$-algebra, with zero differential and zero Lie bracket. Here we
assume that $d\geq2$. The Hochschild cohomology of this algebra is,
as a $\Z$-graded vector space, the algebra of polynomials
$H^*(\Hoch(A))=\C[x^1,\ldots,x^N,y_1,\ldots,y_N]$ in the doubled set
of variables $x^i,y_i$, where the extra generators have degree
$\deg(y_i)=d-q_i$ \cite{kon2}. In general, for the algebra $\CO(M)$ of
regular functions on a smooth $\Z$-graded algebraic supermanifold $M$,
the Hochschild cohomology is given by the algebra of functions on the
total space of the twisted by $[d]$ cotangent bundle to $M$,
$H^*(\Hoch(\CO(M)))=\CO(T^*[d]M)$. The proof goes along the same lines
as the Hochschild-Kostant-Rosenberg theorem, which gives this result
for associative algebras of functions ($d=1$).

When the Lie bracket on the original $d$-algebra is nonzero, this leads to 
a coboundary operator on the above Hochschild cohomology. To find the 
actual Hochschild cohomology one should take the cohomology with respect 
to this coboundary. This coboundary operator is canonically related to the bracket. 
A bracket on the $d$-algebra corresponds to a Poisson structure $\omega^{ij}$ 
of degree $1-d$ on $M$. When we use local coordinates $(x^i,y_i)$ on $T^*[d]M$, 
as in the polynomial algebras above, the coboundary operator is given locally 
by $\omega^{ij}y_j\frac{\del}{\del x^i}$, which indeed has degree 1. We can 
also give this differential operator globally on $T^*[d]M$. We denote the 
pull-back of the Poisson structure $\omega$ to the full space also by $\omega$. 
The total space $T^*[d]M$ has a canonical 1-form $\theta$. This 
1-form is such that the canonical symplectic structure is given by $d\theta$, 
and in local coordinates is given by $\theta=y_idx^i$ (this differential form 
might be familiar from classical mechanics, where it is usually denoted $p_idq^i$). 
Contracting the bi-vector $\omega$ with this form leads to a vector field 
$\theta\cdot\omega$, generating the above differential.

\section{Double Complexes and Spectral Sequences}
\label{app:dcplx}

In this appendix we shortly discuss double complexes and their cohomology. 
For more details see e.g.\ \cite{botu}. 
A double complex consists of a set of vector spaces $C^{p,q}$ carrying two degrees, 
together with two mutually anticommuting coboundary operators $d$ and $\delta$, 
so $d^2=\delta^2=d\delta+\delta d=0$.\footnote{One usually considers commuting 
coboundary operators, introducing extra sign factors in the formulas. It can easily 
seen however that this is completely equivalent.} The operator $\delta$ increases 
the first degree $p$ by one, and the $d$ increases $q$ by one. We can draw this 
double complex in a diagram as in \eqref{dcompl}, with the operator $\delta$ acting 
horizontally and $d$ acting vertically. 
\begin{equation}\label{dcompl}
\matrix{
C^{0,0}     & \labrarr{\delta}& C^{1,0}& \rarr & C^{2,0}& \rarr & C^{3,0}& \rarr & \cdots\cr
 \labdarr{d}&                 & \darr  &       & \darr  &       & \darr  &       &       \cr
C^{0,1}     & \rarr           & C^{1,1}& \rarr & C^{2,1}& \rarr & C^{3,1}& \rarr & \cdots\cr
 \darr      &                 & \darr  &       & \darr  &       & \darr  &       &       \cr
C^{0,2}     & \rarr           & C^{1,2}& \rarr & C^{2,2}& \rarr & C^{3,2}& \rarr & \cdots\cr
 \darr      &                 & \darr  &       & \darr  &       & \darr  &       &       \cr
 \vdots     &                 & \vdots &       & \vdots &       & \vdots &       & 
}
\end{equation}
To any double complex one can canonically connect a complex, where the total degree 
equals the sum of the two degrees, so that the degree $k$ space of this complex 
is given by 
\begin{equation}
C^k=\bigoplus_{p+q=k}C^{p,q}.
\end{equation}
The total coboundary operator on this complex is given by $D = d + \delta$. 
The essential property $D^2=0$ can easily be checked from the analogous property 
of the two coboundary operators. Also, it is clear that it increases the total 
degree $k$ by one. There is now a very convenient way to calculate the total 
cohomology $H^*_D(C)$ of this induced complex. The idea is to calculate separately 
the $d$ and $\delta$ cohomology. First one calculates cohomology with respect to $d$, 
\begin{equation}
E_1 = H_d(C).
\end{equation}
This is the first approximation to the total cohomology. The operator $\delta$ in 
general also induces a coboundary operation on this cohomology, which we also denote 
by $\delta$. We can now make a better approximation of the total cohomology by 
taking the cohomology with respect to this coboundary, 
\begin{equation}
E_2 = H_\delta(E_1). 
\end{equation}
In general however, there may still be a coboundary operator left on the result. 
This procedure can be repeated, leading to a series of complexes $E_r$ with coboundary 
operator $d_r$, 
\begin{equation}
E_r = H_{d_r}(E_{r-1}),
\end{equation}
with the $r$th coboundary operator having degree $(r,1-r)$. 
One usually find that $E_r$ becomes stationary after a certain point. This 
happens for example if the range of one of the bidegrees is finite, so that 
$d_r$ must vanish for sufficiently large $r$.

In the spectral sequence we can represent a class in the zeroth term $E_0$ 
by a $d$-closed element $\al_0$. In the first term $E_1$ we take the 
cohomology with respect to $\delta$, but in the $d$-cohomology. This means 
that $\al_0$ should be $\delta$-closed up to the image of $d$. A class 
in $E_1$ is therefore represented by a pair ($\al_0,\al_1)$, such that 
with $d\al_0=0$, and $\delta\al_0=-d\al_1$. Now in general, 
the second term $E_2$ has a remaining coboundary operator. The coboundary 
operator acting on the representing element $\al_0$ is given by the class 
of $\delta\al_1$, $d_2[\al_0]=[\delta\al_1]$. This can be depicted 
as follows
\begin{equation}
\matrix{
 & & & & & \vdots & & & & \cr
 & & & \al_1 & \rightarrow & d_2\al_0 \cr
 & & & \downarrow \cr
 & \al_0 & \rightarrow & d_1\al_0 \cr
 & \downarrow \cr
 & 0
}
\end{equation}
where $d$ acts vertically and $\delta$ acts horizontally. 
For $\al$ to represent a cohomology class in $E_2$, this requires $d_2\al$ 
to be zero. Remember however that we are still working in the $d$-cohomology, 
therefore it only needs to be zero as a class in this cohomology. In other words, 
it only needs to be zero modulo a $d$-exact term. This repeats the diagram 
above until at some point it terminates, when the 
differential is zero. It gives rise to a sequence of equations, 
\begin{equation}
d\al_0 = 0,\qquad
\delta\al_0 = -d\al_1,\qquad
\delta\al_1 = -d\al_2,\qquad
\delta\al_2 = -d\al_3,\qquad
\cdots.
\end{equation}
These are the same as the familiar descent equations. 
It is easily checked that the total representative $\al=\al_0+\al_1+\cdots$ 
is closed with respect to total coboundary $D$.


\begin{thebibliography}{88}

\bibitem{codo} A.~Connes, M.R.~Douglas and A.~Schwarz,
\textit{Noncommutative Geometry and Matrix Theory: Compactification on Tori}, 
\jhep{02}{1998}{003},
\hepth{9711162}. 

\bibitem{dohu} M.R.~Douglas and C.~Hull, 
\textit{D-Branes and the Noncommutative Torus}, 
\jhep{02}{1998}{008}, 
\hepth{9711165}.

\bibitem{hove} C.M.~Hofman and E.~Verlinde,
\textit{U-Duality of Born-Infeld on the Noncommutative Two-Torus}, 
\jhep{12}{1998}{010}, 
\hepth{9810116}.

\bibitem{scho} V.~Schomerus,
\textit{D-branes and Deformation Quantisation},
\jhep{06}{1999}{030}, 
\hepth{9903205}.

\bibitem{seiwit} N.~Seiberg and E.~Witten, 
\textit{String Theory and Noncommutative Geometry}, 
\jhep{09}{1999}{032},
\hepth{9908142}.

\bibitem{kon1} M.~Kontsevich,
\textit{Deformation Quantisation of Poisson Manifolds, I},
\texttt{math.QA/9709180}.

\bibitem{kon2} M.~Kontsevich,
\textit{Operads and Motives in Deformation Quantisation},
\texttt{math.QA/9904055}.
\hepth{9711162}.

\bibitem{tam} D.~Tamarkin, 
\textit{Another Proof of M.~Kontsevich Formality Theorem},
\texttt{math/9803025}.

\bibitem{kon3} M.~Kontsevich, Y.~Soibelman,
\textit{Deformations of Algebras over Operads and Deligne's Conjecture}
\texttt{math.QA/0001151}

\bibitem{cafe} A.S.~Cattaneo and G.~Felder, 
\textit{A Path Integral Approach to the Kontsevich Quantisation Formula},
\texttt{math.QA/9902090}.

\bibitem{homaOS} C.M.~Hofman and W.K.~Ma,
\textit{Deformations of Topological Open Strings}, 
\jhep{0101}{2001}{035}, 
\hepth{0006120}. 

\bibitem{laza} C.I.~Lazaroiu,
\textit{On the Structure of Open-Closed Topological Field Theory in Two Dimensions},
\hepth{0010269}. 

\bibitem{wizwi} E.~Witten and B.~Zwiebach,
\textit{Algebraic Structures and Differential Geometry in 2D String Theory},
\npb{377}{1992}{55-112},
\hepth{9201056}.

\bibitem{zwie} B.~Zwiebach,
\textit{Closed String Field theory: Quantum Action and the BV Master Equation}, 
\npb{390}{1993}{33-152},
\hepth{9206084}.

\bibitem{stasheff} J.~Stasheff,
\textit{Closed String Field Theory, Strong Homotopy Lie Algebras and 
the Operad Actions of Moduli Space},
\hepth{9304061}

\bibitem{kvz} T.~Kimura, A.A.~Voronov, G.J.~Zuckerman,
\textit{Homotopy Gerstenhaber Algebras and Topological Field Theory},
\texttt{q-alg/9602009}.

\bibitem{tam2} D.~Tamarkin, 
\textit{The Deformation Complex of a $d$-Algebra is a $(d+1)$-Algebra},
\texttt{math/0010072}.

\bibitem{bebe} E.~Bergshoeff, D.S.~Berman, J.P.~van der Schaar, and P.~Sundell
\textit{A Noncommutative M-Theory Five-Brane},
\hepth{0005026}.

\bibitem{kasa} S.~Kawamoto and N.~Sasakura,
\textit{Open Membranes in a Constant C-field Background and Noncommutative 
Boundary Strings}, 
\hepth{0005123}.

\bibitem{matshi} Y.~Matsuo and Y.~Shibusa,
\textit{Volume Preserving Diffeomorphism and Noncommutative Branes},
\jhep{02}{2001}{006}, 
\hepth{0010040}, 

\bibitem{lomosh} A.~Losev, G.~Moore, and S.L.~Shatasvili,
\textit{M\&m's},
\npb{522}{1998}{105}, 
\hepth{9707250}.

\bibitem{aha} O.~Aharony,
\textit{A Brief Review of ``Little String Theories''}, 
\cqg{17}{2000}{929-938},
\hepth{9911147}.

\bibitem{om} R.~Gopakumar, S.~Minwalla, N.~Seiberg and A.~Strominger,
\textit{OM Theory in Diverse Dimensions},
\hepth{0006062}.

\bibitem{bebe2} E.~Bergshoeff, D.S.~Berman, J.P.~van der Schaar, and P.~Sundell,
\textit{Critical Fields on the M5-Brane and Noncommutative Open Strings},
\plb{492}{2000}{193}, 
\hepth{0006112}.

\bibitem{witeqn} E.~Witten,
\textit{Two-Dimensional Gravity and Intersection Theory on Moduli Space},
Surv.\ Diff.\ Geom.\ \textbf{1} (1991) 243-31.

\bibitem{dvv} R.~Dijkgraaf, E.~Verlinde and H.~Verlinde,
\textit{Topological Open Strings in $d<1$},
\npb{352}{1991}{59-86}.

\bibitem{js} J.-S.~Park,
\textit{Topological Open $p$-Branes}, 
\hepth{0012141}.

\bibitem{stas} J.~Stasheff,
\textit{Homotopy Associativity of H-Spaces}, I and II,
Trans.\ Amer\. Math\. Soc.\ \textbf{108} (1963) 275 and 293.

\bibitem{getzjon1} E.~Getzler and J.D.S.~Jones, 
\textit{$A_\infty$-Algebras and the Cyclic Bar Complex},
Ill.\ J.\ Math.\ \textbf{34}, No.~2 (1990) 256.

\bibitem{getzjon2} E.~Getzler and J.D.S.~Jones, 
\textit{Operads, Homotopy Algebra, and Interated Integrals for Double Loop Spaces},
\hepth{9403055}.

\bibitem{zwieoc} B.~Zwiebach, 
\textit{Oriented Open-Closed String Theory Revisited},
\ap{267}{1998}{193-248},
\hepth{9705241}.

\bibitem{kon4} M.~Kontsevich,
\textit{Feynman Diagrams and Low-Dimensional Topology}, 
First European Congress of Mathematics, 1990-1992 
Birkh\"auser (1993) 173.

\bibitem{ksv} T.~Kimura, J.~Stasheff, A.A.~Voronov,
\textit{On Operad Structures of Moduli Spaces and String Theory},
\hepth{9307114}.

\bibitem{liz} B.H.~Lian and G.J.~Zuckerman,
\textit{Algebraic and Geometric Structures in String Backgrounds},
\hepth{9506210}.

\bibitem{tamtsy} D.~Tamarkin and B.~L.~Tsygan,
\textit{Noncommutative Differential Calculus, Homotopy BV Algebras and Formality Conjectures},
\texttt{math.KT/0002116}.

\bibitem{verl} E.~Verlinde,
\textit{The Master Equation of 2d String Theory},
\npb{381}{1992}{141-157},
\hepth{9202021}.

\bibitem{tom} C.~Hofman, J.-S.~Park, and S.-M.~Lee,
to be published.

\bibitem{witcs} E.~Witten,
\textit{Chern-Simons Gauge Theory as a String Theory},
\hepth{9207094}.

\bibitem{cour} T.~Courant,
\textit{Dirac Manifolds}, 
Trans.\ Amer.\ Math.\ Soc.\ \textbf{319} (1990) 631, 
\texttt{funct-an/9702004}.

\bibitem{royt} D.~Roytenberg,
\textit{Courant Algebroids, Derived Brackets and Even Symplectic Supermanifolds},
Ph.D. thesis, University of California at Berekeley (1999)
\texttt{math.DG/9910078}.

\bibitem{wein} Z.-J.~Liu, A.~Weinstein, and P.~Xu,
\textit{Manin Triples for Lie Bialgebroids}, 
J.\ Diff.\ Geom.\ \textbf{45} (1997) 547, 
\texttt{dg-ga/9508013}.

\bibitem{hofpar} C.~Hofman and J.-S.~Park,
work in progress.

\bibitem{botu} R.~Bott and L.W.~Tu, 
\textit{Differential Forms in Algebraic Topology},
Springer-Verlag, New York, 1982.


\end{thebibliography}
\end{document}